# Spontaneous Surface Charging and Janus Nature of the Hexagonal Boron Nitride-Water Interface

Yongkang Wang[1,*,#], Haojian Luo[1#], Xavier R. Advincula[2,3,4#], Zhengpu Zhao[5#], Ali Esfandiar[1,6], Da Wu[5], Kara D. Fong[2,4], Lei Gao[1], Arsh S. Hazrah[1], Takashi Taniguchi[7], Christoph Schran[3,4], Yuki Nagata[1], Lydéric Bocquet[6], Marie-Laure Bocquet[6], Ying Jiang[5], Angelos Michaelides[2,4], Mischa Bonn[1*]

[1] Department of molecular spectroscopy, Max Planck Institute for Polymer Research, Ackermannweg 10, 55128 Mainz, Germany.

[2] Yusuf Hamied Department of Chemistry, University of Cambridge, Lensfield Road, Cambridge, CB2 1EW, UK.

[3] Cavendish Laboratory, Department of Physics, University of Cambridge, Cambridge, CB3 0HE, UK.

[4] Lennard-Jones Centre, University of Cambridge, Trinity Ln, Cambridge, CB2 1TN, UK.

[5] International Center for Quantum Materials, School of Physics, Peking University, Beijing 100871, China.

[6] Laboratoire de Physique de l'École Normale Supérieure, ENS, Université PSL, CNRS, Sorbonne Université, Université Paris Cité, F-75005 Paris, France.

[7] Research Center for Materials Nanoarchitectonics, National Institute for Materials Science, Tsukuba, Japan.

[#]Those authors contributed equally to this work.

[*]Correspondence to: wangy3@mpip-mainz.mpg.de, bonn@mpip-mainz.mpg.de

## Abstract

Boron, nitrogen and carbon are neighbors in the periodic table and can form strikingly similar twin structures—hexagonal boron nitride (hBN) and graphene—yet nanofluidic experiments demonstrate drastically different water friction on them. We investigate this discrepancy by probing the interfacial water and atomic-scale properties of hBN using surface-specific vibrational spectroscopy, atomic-resolution atomic force microscopy (AFM), and machine learning-based molecular dynamics. Spectroscopy reveals that pristine hBN acquires significant negative charges upon contacting water at neutral pH, unlike hydrophobic graphene, leading to interfacial water alignment and stronger hydrogen bonding. AFM supports that this charging is not defect-induced. pH-dependent measurements suggest $OH^−$ chemisorption and physisorption, which simulations validate as two nearly equally stable states undergoing dynamic exchange. These findings challenge the notion of hBN as chemically inert and hydrophobic, revealing its spontaneous surface charging and Janus nature, and providing molecular insights into its higher water friction compared to carbon surfaces.

# Main Text

Water transport at the nanoscale plays a crucial role in numerous biological and industrial processes, from neurotransmission to ultrafiltration[1,2], thus attracting substantial interest. Recent advances in nanofluidics have enabled the development of artificial nanochannels with dimensions as small as a few angstroms, using atomically smooth surfaces like one-dimensional (1D) channels formed by carbon and boron nitride nanotubes[3,4], as well as two-dimensional (2D) channels made from 2D materials such as graphene and hexagonal Boron Nitride (hBN)[5–8]. These developments have facilitated a deeper exploration of water transport properties at the nanoscale, uncovering unexpected and significant differences in water's hydrodynamic friction on those atomically smooth surfaces, with hBN consistently exhibiting one to two orders of magnitude higher friction than graphene, whether quantified by mass transfer[6], boundary slip length[4,9] or friction coefficient[7,10]. While current theories of solid-liquid interfaces typically describe the solid as a static external potential that influences the behavior of fluid molecules, with friction primarily attributed to the solid's surface roughness[11], only a three to five times difference in water's hydrodynamic friction is expected[12,13] given that hBN and graphene share similar allotropic forms, which are often considered comparable in terms of surface roughness and presumed hydrophobicity[14,15]. Indeed, several additional anomalous phenomena/properties of water have been observed in hBN nanoconfinement, such as spontaneous hydrolysis[16], osmotic energy conversion[17], atypical aqueous ion transport[18], and giant ferroelectric-like in-plane dielectric constant and notably enhanced in-plane conductivity[19].

These observations point to an unexpectedly strong interaction of water with hBN, but the underlying mechanism has remained elusive or controversial. Previous studies have relied primarily on theoretical and computational simulations[12–14,20–24], while experimental insights remain scarce. Remarkably, nanofluidics experiments have indicated that surface charging for hBN in contact with water may serve as a possible explanation[6,8,17,18,25,26]. While surface charges would indeed substantially enhance the interaction of hBN with water, the possible origin of the charge remains debated. For instance, while atomically flat 2D materials are typically considered charge-neutral and hydrophobic[14,27], theoretical studies have suggested that hydroxide (OH$^-$) ions, a product of water autoionization, may exhibit an affinity for the hBN surface[16,23,24]. This indicates that the hBN surface might undergo surface charging through the adsorption of OH$^-$ when interacting with water. Such external surface charging could impact water transport by enhancing electrostatic interactions[24,28] and by roughening locally the flat sheet. Furthermore, it has been proposed that defects on the hBN surface, often inevitable during crystal growth, may influence water transport similarly to the charging effect[20–22]. Both mechanisms provide plausible explanations for the differences in water transport behavior between hBN and graphene and challenge the notion of hBN's "chemical inertness." Yet, while the defect scenario is extrinsic and could potentially be mitigated by treatment or using improved hBN, the adsorption of OH$^-$ ions sets an intrinsic limitation on hBN's properties for nanofluidics.

Clearly, molecular-level insights into the potential occurrence of surface charging at 2D materials (hBN and graphene)-water interfaces are essential to verify or falsify these mechanisms. Ideally, one would like access to the molecular-level details of the buried 2D material-water interface, including interfacial water structure, such as its orientation and hydrogen bond (H-bond) environment, as well as the hBN surface properties like defects and surface charges. Here, we provide such molecular-level insights by combining heterodyne-detected sum frequency generation (HD-SFG) spectroscopy, atomic-resolution atomic force microscopy (AFM), and machine learning-based molecular dynamics.

HD-SFG spectroscopy is an ideal tool for investigating the interfacial water structure and the potential presence of surface charges on 2D materials. As a surface-specific vibrational spectroscopy technique, HD-SFG selectively probes the water molecules at the interface[14,29,30], naturally excluding signals from bulk water due to the SFG selection rules[31,32]. This method provides access to the complex $\chi^{(2)}$ spectrum, where the imaginary part ($\text{Im}(\chi^{(2)})$) reveals crucial information about the H-bond network and the absolute orientation of interfacial water molecules[33,34]. Moreover, the additive nature of the $\chi^{(2)}$ signals allows for the separation of contributions from water aligned as the result of surface charge, enabling the direct quantification of surface charges[35,36]. These capabilities make HD-SFG spectroscopy particularly well-suited for probing interfacial water and surface charge information on the hBN, offering new experimental insights into its "chemical inertness."

In addition to HD-SFG spectroscopy to examine the interfacial water structure and surface charges on hBN, we also use AFM to visualize surface defects in real space. Our method involves the preparation of high-quality, single-crystal hBN via mechanical exfoliation, yielding a defect-free hBN surface. The qPlus-based AFM measurements confirm the absence of defects, while the HD-SFG spectroscopy reveals that the interfacial water molecules form strong hydrogen bonds and are aligned up toward the defect-free, negatively charged hBN. Interestingly, the defect-free hBN surface exhibits significant negative charging when in contact with water, even at neutral pH, unlike graphene, which we show remains charge-neutral and hydrophobic under similar conditions. We attribute this surface charging to the adsorption of $OH^-$ ions on the hBN surface, supported by pH-dependent surface charge measurements from HD-SFG spectroscopy. Additionally, through machine learning-based molecular dynamics simulations with first-principles accuracy, we demonstrate that $OH^-$ adsorption occurs in two almost equally stable states—chemisorbed and physisorbed. These states are also separated by a low energy barrier, facilitating dynamic interconversion between them. Our experimental results and atomistic simulations challenge the traditional view of hBN as "chemically inert", and offer new insights into the mechanisms behind surface charging in two-dimensional materials.

We prepared a large-area (> 200 × 200 μm²) hBN flake, approximately 100 nm thick, on a $SiO_2$ substrate using the well-established polymer and solvent-free mechanical exfoliation method following the procedures described in the refs[6,7]. The procedures are detailed in the Methods. After flake preparation, a flat and clean region approximately 150 × 150 μm² in size was identified using an optical microscope, and a 100 nm thick gold film was used to mark the identified area and encapsulate/cover the edges of the hBN for the HD-SFG measurement (See Supplementary Methods S1-S3 for more details). We ensured the selected region was clean and atomically smooth, free of visible wrinkles, edges, and hydrocarbon contamination (see Supplementary Note S1 for details). Additionally, we confirmed that the ~100 nm thick hBN layer effectively screened any potential influence of the supporting substrate on the interfacial water at the supported hBN/water interface (see Supplementary Note S2 and Note S3 for details). A schematic of the sample composition and beam geometry of SFG measurement is shown in Fig. 1a, and an optical image of the prepared hBN sample from the bottom view is shown in Fig. 1b. In this HD-SFG configuration, the visible ($\omega_{vis}$), local oscillator (LO), and infrared ($\omega_{IR}$) lights impinge non-collinearly from the optically transparent $SiO_2$ substrate, passing through the $SiO_2$ and the hBN flake, to overlap at the hBN/water interface. The reflected LO light interferes with the sum-frequency ($\omega_{SFG}$) signals generated from the water in the 'reflected' direction, producing a heterodyned sum-frequency output that enables access to the $Im(\chi^{(2)})$ signals.

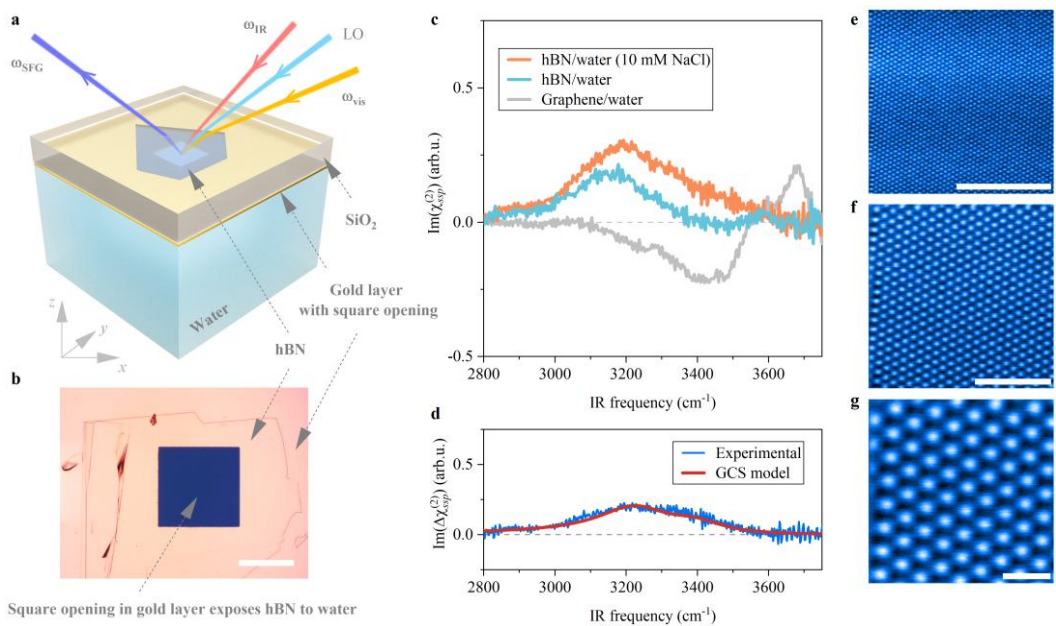

**Fig. 1 | Interfacial water structure on hBN revealed by HD-SFG spectroscopy**. **a**. A schematic of the composition of the hBN sample and beam geometry of the SFG setup. The hBN flake is positioned on an $SiO_2$ substrate, with a gold layer encapsulating its edges. A square opening in the gold layer, located near the center of the hBN, exposes a portion of the hBN surface to water. The

laser beams reach the hBN/water interface through the substrate. **b**. An optical image of the prepared hBN sample on a SiO$_2$ substrate, encapsulated by gold along its edges, showing the bottom view of the setup described in (**a**). The scale bar corresponds to 100 μm. **c**. Experimental $\text{Im}(\chi_{\text{BN}}^{(2)})$ spectrum obtained for water and 10 mM NaCl at pH~6. Experimental $\text{Im}(\chi_{\text{G}}^{(2)})$ spectrum of the graphene-water interface is shown for comparison. **d**. Experimental difference spectrum $\text{Im}(\Delta\chi_{\text{BN}}^{(2)})$ between 10 mM and 100 mM NaCl solutions, compared with a calculated spectrum based on the Gouy-Chapman-Stern (GCS) theory. The grey dashed lines in (**c**) and (**d**) represent zero lines. **e**. Constant-height AFM image of the hBN surface. **f** and **g**. Zoomed-in AFM images from (**e**), with B and N atoms depicted in white and black, respectively. The scale bars indicate 5 nm, 2 nm, and 0.5 nm, respectively. SFG, sum-frequency generation light; vis, visible light; IR, infrared light; ω, angular frequency of light; LO, local oscillator; arb.u., arbitrary units.

We conducted the HD-SFG measurement within the marked region on the hBN sample using our homemade flow cell at the *ssp* polarization combination with the three letters indicating the polarizations of the SFG, visible, and infrared light fields, respectively (Fig. 1a, see Supplementary Methods S4 for more details). The $\text{Im}(\chi_{\text{BN}}^{(2)})$ spectrum of water in contact with hBN measured in the 2800-3750 cm$^{-1}$ frequency region using pure water (pH ~6) is displayed in Fig. 1c. This spectrum exhibits primarily a broad positive O-H stretch peak spanning from 2900 cm$^{-1}$ to 3500 cm$^{-1}$. The positive sign of the peak indicates the O-H group of the interfacial water pointing up[37] towards the hBN surface, and its low peak frequencies indicate that the O-H group forms strong H-bonds[14,38]. The $\text{Im}(\chi_{\text{BN}}^{(2)})$ spectrum contrasts sharply with the $\text{Im}(\chi_{\text{G}}^{(2)})$ spectrum measured at the graphene/water interface, as shown in Fig. 1d. The $\text{Im}(\chi_{\text{G}}^{(2)})$ spectrum closely resembles that of a hydrophobic interface, such as the air/water interface[31,32,39] and alkane/water interface[40–42], featuring a broad negative hydrogen-bonded (H-bonded) O-H peak around 3400 cm$^{-1}$ and a positive high-frequency dangling O-H peak above 3600 cm$^{-1}$, originating from OH groups pointing up towards graphene[14,27,43,44]. This suggests that the graphene surface is hydrophobic and chemically inert in contact with water, consistent with previous experimental measurements[27,45] and theoretical predictions[14,43]. Interestingly, earlier theoretical studies employing *ab initio* molecular dynamics (AIMD) predicted that the pristine hBN surface would likewise be hydrophobic, with an $\text{Im}(\chi_{\text{BN}}^{(2)})$ spectrum similar to $\text{Im}(\chi_{\text{G}}^{(2)})$, showing a broad negative H-bonded O-H peak around 3400 cm$^{-1}$ and a positive high-frequency dangling O-H peak above 3600 cm$^{-1}$[14]. However, our experimental $\text{Im}(\chi_{\text{BN}}^{(2)})$ spectrum reveals only a positively signed H-bonded O-H peak at a low frequency (~3150 cm$^{-1}$), with no noticeable signature of the dangling O-H peak. Our finding implies that the hBN surface is not hydrophobic but hydrophilic and negatively charged when in contact with water at neutral pH. Notably, the absence of C-H peaks (2850–2950 cm$^{-1}$) in these $\text{Im}(\chi_{\text{BN}}^{(2)})$ spectra underscores the cleanliness of the samples, free of hydrocarbon contamination[46,47]. We also checked that the observed spectrum features do not arise from carbonate in the water (see Supplementary Note S4 for details).

To further support that the hBN surface is negatively charged upon contacting water, we measured the $\text{Im}(\chi_{\text{BN}}^{(2)})$ spectrum upon adding 10 mM NaCl to the water. At a charged interface, in addition to the surface contribution ($\chi_{\text{s}}^{(2)}$) arising mainly from the alignment of the topmost 1-2 layers of water, the penetration of the electrostatic field into the bulk solution induces alignment and polarization of water molecules in the diffuse layer, providing a bulk contribution ($\chi^{(3)}$) to the SFG signals[36,48–50], *i.e.*, $\chi^{(2)}(\sigma_0, c) = \chi_{\text{s}}^{(2)}(\sigma_0) + \chi^{(3)}(\sigma_0, c)$ (See Supplementary Note S5 for more details). The addition of electrolyte with concentration $c$ screens the surface charge ($\sigma_0$), which in turn modifies the bulk $\chi^{(3)}$ contribution to the SFG spectrum that scales with surface potential[36,48–50]. This bulk contribution, if present, should be significantly modified, while the surface $\chi_{\text{s}}^{(2)}$ contribution remains weakly affected[36]. The data, shown in Fig. 1c, reveals a substantial modification of the water response, indicating the hBN surface is indeed charged. A quantitative analysis of the differential SFG signals at different ion strengths, confirms a significant bulk $\chi^{(3)}$ contribution peaked at around 3250 cm$^{-1}$[36,48–50], whose positive sign further confirms the surface's negative charge (Fig. 1d). Following previous protocols within the Gouy-Chapman-Stern (GCS) electric double-layer (EDL) model[36,48–50], we infer from the differential SFG signal that $\sigma_0$ on hBN at pH~6 is -15±6 mC/m$^2$ (See Supplementary Note S5 for more extensive discussion on this estimation).

What is the mechanism behind the surface charging of the hBN? The hBN surface may acquire negative charges upon contact with pure water mainly for two possible reasons: (i) the presence of defects such as boron vacancies[20,25], and (ii) the adsorption of $OH^-$ ions[24], a product of water autoionization ($H_2O \leftrightarrow OH^- + H^+$), on the hBN surface. To examine (i) the potential presence of defects on the hBN surface, we conducted qPlus-based AFM measurements. All AFM data were acquired at 6 K under ultra-high vacuum conditions (<5×10$^{-10}$ Torr) to probe potential atomic defects. The constant-height, high-resolution AFM images of the hBN surface from a randomly selected region, shown in Fig. 1e-g, reveal a clean surface with a perfect hexagonal honeycomb structure without defects over an area of 100 nm². We conducted the qPlus-based AFM measurements at five different randomly selected 100 nm² regions and all data show the absence of defects on the hBN surface (See Supplementary Note S6 for more results). The estimated surface charge density on hBN of -15 mC/m$^2$ corresponds to one charge per ~11 nm$^2$. The probability of not finding a defect at this density across five different areas of 100 nm² is below ~5×10$^{-21}$, assuming Poisson distribution of defects. We therefore conclude that defects are not the primary cause of the surface charging observed on the hBN surface.

The above analysis indicates that defects are not responsible and implies that the adsorption of $OH^-$ ions on the hBN surface might be responsible for the negative surface charge. The hypothesis of adsorption of $OH^-$ ions on the hBN surface is plausible, given the appearance of the positive peak with a low peak frequency at approximately 3150 cm$^{-1}$ in the $Im(\chi^{(2)}_{BN})$ spectrum (Fig. 1c). The peak frequency of 3150 cm$^{-1}$ is about 100 cm$^{-1}$ red-shifted compared to the bulk $\chi^{(3)}$ contribution (Fig. 1d), which peaks at around 3250 cm$^{-1}$ regardless of salt solution or surface properties[36,48–50]. This redshift can be accounted for by interfacial water O-H groups donating strong H-bonds to OH$^-$ at the hBN interface. Remarkably, the 3150 cm$^{-1}$ peak exhibits a continuum extending below 2900 cm$^{-1}$, testifying to the strong interaction of water O-H groups with OH$^-$ species[51]. These water O-H groups, on average, point *up* towards the adsorbed OH$^-$ on the hBN surface, which explains its positive sign.

These experimental findings strongly indicate that OH$^-$ ions adsorb at the hBN interface, influencing the orientation of interfacial water molecules. The absence of a strong chemisorbed O-H signature in the $\text{Im}(\chi_{\text{BN}}^{(2)})$ spectrum, which would feature a negative high-frequency peak around 3600–3670 cm$^{-1}$ such as observed on CaF$_2$[52], sapphire[53], and silica surfaces[54], indicates a more complex adsorption behavior on hBN, possibly (also) involving physisorption rather than purely strong covalent bonding through chemisorption. Given the unexpected surface charging and the distinct spectral features observed, a deeper understanding of the underlying adsorption mechanisms is needed.

Motivated by these experimental observations, we conducted machine learning-based molecular dynamics (MD) simulations with first-principles accuracy for liquid water films at hBN interfaces (see Methods). Specifically, we investigated where OH$^-$ ions adsorb at the interface and how they interact with water through a series of constrained and free MD simulations. A key result of this analysis is shown in Fig. 2a where we report the potential of mean force (PMF) of an OH$^-$ ion as a function of its distance from hBN. These simulations reveal two stable adsorption states on hBN, illustrated in Fig. 2b. The first is a well-defined chemisorbed state with the OH$^-$ covalently bonded to a boron atom of the hBN layer at approximately 1.6 Å. The second, which we refer to as the physisorbed state, has the OH$^-$ solvated within the first contact layer of water at around 3.4 Å from the surface. The stabilities of the two states are similar, with a free energy (relative to an OH$^-$ in the interior of the water film) of 0.09 ± 0.02 eV for the chemisorbed state and -0.02 ± 0.01 eV for the physisorbed state. The presence of two states is consistent with a previous AIMD study[24]. However, the behavior seen here on hBN is in stark contrast to graphene, where only a physisorbed state is observed at approximately 3.4 Å[55], highlighting a key difference between the two materials.

Our free energy calculations show that the barrier between the chemisorbed and physisorbed states is low; approximately 0.2 eV to go from the chemisorbed to the physisorbed state. This low barrier, along with the similar free energy of the two states, suggests the possibility of dynamic exchanges between these two configurations, indicating a more intricate adsorption behavior than previously recognized[24]. This finding points to an intriguing surface charging scenario involving both static (chemisorbed) and dynamic (physisorbed) surface charges. Indeed, upon running free MD, we see transitions from the chemisorbed to the physisorbed state on the nanosecond timescale, consistent with the barrier obtained from constrained MD (see Supplementary Note S7). A closer inspection of the free MD simulations reveals an interesting transition mechanism: the chemisorbed $OH^-$ first undergoes protonation before desorbing as a water molecule. This process is illustrated schematically in Fig. 2c and is visible in Supplementary Movie 1. Additionally, we examined the dynamics of the two states and found clear differences. In the chemisorbed state, $OH^-$ remains relatively immobile, tightly bound to boron, while in the physisorbed state, it gains in-plane mobility, allowing freer diffusion along the surface (see Supplementary Note S7). This distinction is particularly relevant for understanding nanoscale friction on hBN, as the mobility of surface-bound species can significantly influence interfacial slip and energy dissipation.

We now examine how the $OH^-$ ion impacts the surrounding water in its chemisorbed and physisorbed states. Beyond their energetic similarities, these adsorption states exhibit distinct orientations along the surface normal, directly influencing the alignment of interfacial water molecules (see Fig. 2d). Specifically, when the $OH^-$ is in the physisorbed state the liquid water structure is similar to that of neutral water without any hydroxide. A similar effect is observed when $H_3O^+$ is present at the interface, where the surrounding water molecules retain their neutral water orientational distribution. In contrast, when the OH⁻ is chemisorbed, the hydrogen-bonded network of water is more structured with a peak in the orientational distribution at $\cos\theta \approx -0.5$, corresponding to a preponderance of water molecules oriented towards the surface. This distinction was less evident in a previous AIMD study[14] due to the limited time scales sampled (see Supplementary Note S7 for further details). This again highlights the key role of machine learning-based MD simulations in enabling robust conclusions to be drawn from well-converged simulations.

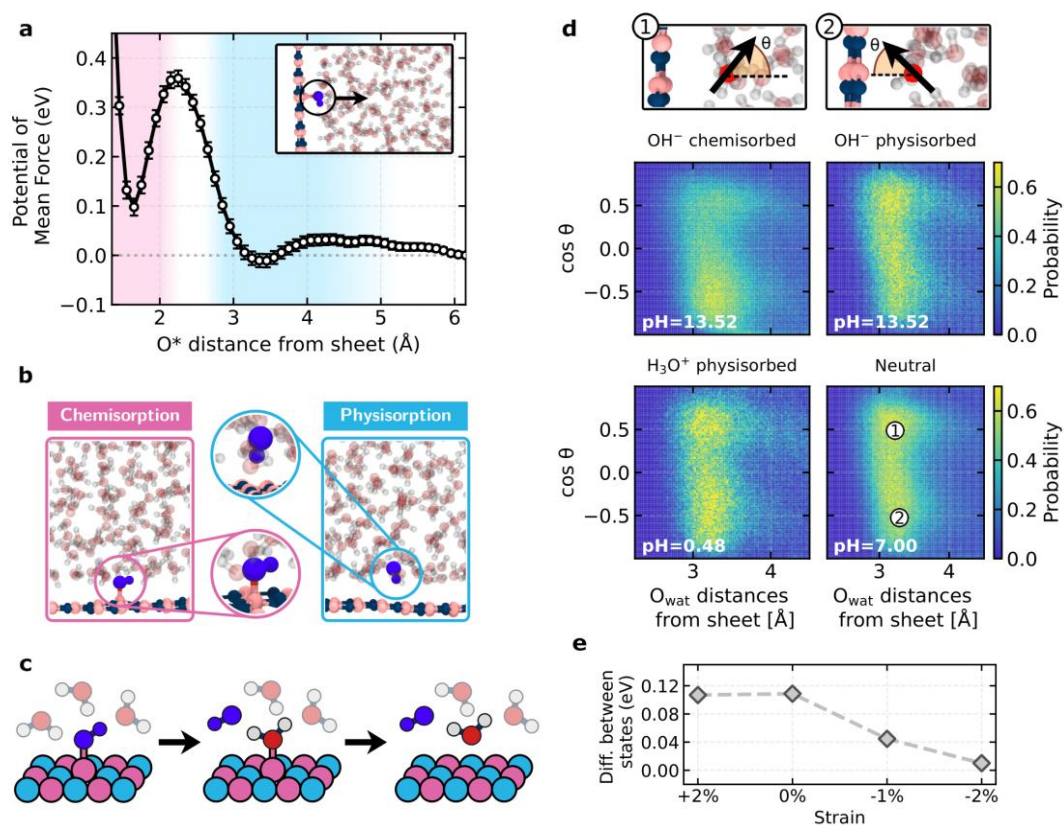

**Fig. 2 | Surface chemistry of hBN revealed by machine learning-based MD simulations**. **a.** Potential of mean force of an OH⁻ ion as a function of its oxygen distance from the hBN, obtained via umbrella sampling. **b.** Representative snapshots of the chemisorbed and physisorbed states, highlighting structural differences. **c.** Transition mechanism illustrating the protonation of chemisorbed OH⁻, followed by its desorption as a water molecule. **d.** Orientational distributions of interfacial water molecules under different pH conditions (basic, acidic, and neutral), showing distinct alignment patterns for chemisorbed and physisorbed OH⁻ ions. The angle definitions are

shown in the accompanying schematics above. **e.** Free energy difference between the chemisorbed and physisorbed states as a function of strain applied to the hBN surface, indicating how mechanical strain influences the relative stability of these adsorption states. A positive value indicates that the physisorbed state is more stable.

Our machine learning based simulations reveal similar stabilities of the two states. Indeed, a 0.1 eV difference between the two states could easily be within the simulation error bar for a complex system such as this. For example, simulations of water are known to be sensitive to nuclear quantum effects and/or different exchange-correlation functionals[56,57]. With this in mind, simulations reported in Supplementary Note S7 show that these effects do slightly alter the relative stabilities of the two states. However, the key conclusion – that both states have a similar energy – is not altered. In addition, we show in Fig. 2e that the relative stability of these adsorption states can be modulated by applying uniaxial strain to the hBN surface. This suggests an additional degree of control over OH$^-$ adsorption, where external mechanical effects could shift the balance between chemisorption and physisorption. This observation should be relevant to the behavior of water in intrinsically strained hBN nanotubes.

To further investigate the adsorption of OH$^-$ ions on the hBN surface and the resulting surface charging behavior, we measured the $\text{Im}(\chi_{\text{BN}}^{(2)})$ spectra while varying the OH$^-$ ion concentration (pH). The ionic strength was maintained at 100 mM by adding NaCl to minimize bulk $\chi^{(3)}$ contributions, as shown in Fig. 3a. The 3150 cm$^{-1}$ peak in the $\text{Im}(\chi_{\text{BN}}^{(2)})$ spectrum increases steadily as the pH increases from 4.5 to 11 but disappears at pH below 4.5. Simultaneously, the bulk contribution follows a similar trend with pH change: it is positive at pH values above 4.5 and negative below 4.5, with its intensity increasing at both higher and lower pH values. By comparing SFG spectra at different ion strengths[36,48–50], we infer that $\sigma_0$ varies from +11 mC/m² to -42 mC/m² between pH=3 and 11, reaching a minimum of approximately -0.5 mC/m² at pH=4.5 (Fig. 3b). These results indicate that the isoelectric point of the hBN surface is around pH=4.5, consistent with previous studies[8,17,23]. Importantly, the consistent change of the 3150 cm⁻¹ peak and $\sigma_0$ further supports our assignment of the 3150 cm⁻¹ peak to the O-H group of the topmost layer of water interacting with the adsorbed OH⁻ on the hBN surface.

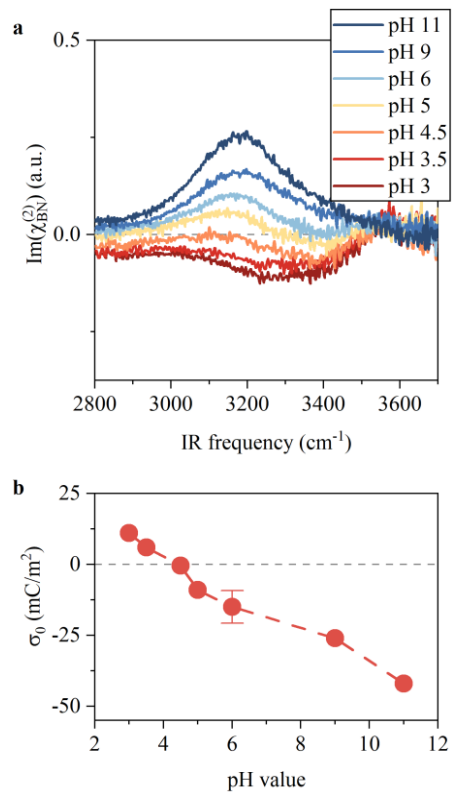

**Fig. 3 | Surface chemistry of hBN revealed by HD-SFG spectroscopy**. **a**. Experimental $\mathrm{Im}\left(\chi_{\mathrm{BN}}^{(2)}\right)$ spectra obtained for 100 mM NaCl at various pH values. **b**. Inferred $\sigma_0$ from HD-SFG signals at various pH values. The grey dashed lines in (**a**) and (**b**) serve as zero lines.

Interestingly, at pH values below 4.5, the water at the interface responds like the hBN surface has become positively charged, and the $\mathrm{Im}\left(\chi_{\mathrm{BN}}^{(2)}\right)$ spectra display a broad H-bonded O-H peak centered around 3350 cm⁻¹, as seen in Fig. 3a. This can be accounted for by protons residing on the topmost layer, contributing to the positively charged interface. The surface propensity of protons has been previously confirmed both experimentally and theoretically for the air/water interface[51,58,59] and the graphene/water interface[55,60], and is also consistent with our simulations shown in Fig. 2d. Our experimental results suggest that the strong surface affinity of protons on hBN is already apparent at low proton concentrations (~0.3 mM, pH=3.5). We tentatively attribute this to the strong affinity of protons for the nitrogen atoms on the hBN surface, analogous to the strong affinity of hydroxide ions for the boron atoms. Regardless, the pH-dependent changes in the $\text{Im}\left(\chi_{\text{BN}}^{(2)}\right)$ spectra provide compelling evidence that challenges the picture of hBN being "chemically inert" when in contact with water. Instead, these results reveal a strong affinity of both OH⁻ ions and protons for the hBN surface, giving rise to a negatively charged interface under mildly basic conditions and a positively charged interface under mildly acidic conditions.

## Conclusions

Our combined experimental and theoretical study challenges the traditional view of hBN's "chemical inertness." Contrary to conventional expectations of a hydrophobic surface, a defect-free hBN surface exhibits substantial negative charging when in contact with water at neutral pH, unlike graphene, which remains charge-neutral and hydrophobic. We provide experimental evidence that this surface charging in hBN arises from the adsorption of OH⁻ ions, a product of water autoionization, aligning with recent theoretical predictions[23,24]. Remarkably, our experimental results suggest that the negative surface charge on hBN is already present under mildly acidic conditions (pH 4.5, OH⁻ concentration of ~3 × 10⁻¹⁰ M) and increases significantly as the pH rises and changes into positive at pH below 4.5. These findings offer molecular-level insights into surface charging mechanisms, prompting a reevaluation of hBN's chemical behavior and intrinsic hydrophilicity. Using machine learning-based molecular dynamics simulations with first-principles accuracy, we further reveal that OH⁻ adsorption occurs in two states—chemisorbed and physisorbed—separated by a low energy barrier, allowing dynamic interconversion between them. This revised understanding may also explain the observed differences in water friction between carbon and hBN surfaces, highlighting the role of surface charging in these variations. Moreover, the inevitable pronounced surface charging due to OH⁻ ion adsorption on defect-free hBN in contact with water at neutral pH should be accounted for when discussing anomalous water properties near hBN surfaces or in nanoscale hBN confinement, such as spontaneous hydrolysis[16], osmotic energy conversion[17], atypical aqueous ions transport[18], and giant ferroelectric-like in-plane dielectric constant and notably enhanced in-plane conductivity[19].

## Methods

***Sample preparation.*** We employed high-quality hBN crystals for sample preparation, obtained from the International Center for Materials Nanoarchitectonics, National Institute for Materials Science 1-1 Namiki, Tsukuba 305-0044, Japan. hBN flakes were mechanically exfoliated using polydimethylsiloxane (PDMS) and dry-transferred onto an oxygen plasma-treated $SiO_2$ substrate. This method ensures clean and large-area sample preparation. After preparation, a flat and clean region approximately $150 \times 150$ µm² in size was identified using an optical microscope, and a gold structure was used to mark the identified area for the HD-SFG measurement. Notably, the thickness of the hBN flake was carefully chosen to be approximately 100 nm, ensuring that the SFG signal primarily probes the hBN/water interface, while minimizing contributions from the $SiO_2$/hBN interface[61]. The preparation of the suspended graphene on the water surface was similar to Refs.[45,62] and was detailed in our recent work[27]. More details of the sample preparation can be found in the Supplementary Method of the Supplementary Information.

***HD-SFG measurement.*** HD-SFG measurements were performed on a non-collinear beam geometry with a Ti:Sapphire regenerative amplifier laser system. A detailed description can be found in refs.[48,52]. HD-SFG spectra were measured in a dried air atmosphere to avoid spectral distortion due to water vapor. To check the sample height, we used a height displacement sensor (CL-3000, Keyence). The IR, visible, and LO beams are directed at the sample (in $SiO_2$) at incidence angles of approximately 34°, 43°, and 41°, respectively. We ensured the power of incident IR (~3 µm) and visible (800 nm) pulses are far below the damage threshold value of a hBN crystal and do not introduce defects on the hBN surface (Supplementary Note S8). The measurements were performed at the *ssp* polarization combination, where *ssp* denotes *s*-polarized SFG,

*s*-polarized visible, and *p*-polarized IR beams. The HD-SFG signal at the hBN/water interface was normalized with that of the hBN/D$_2$O interface. The suspended graphene sample HD-SFG spectra were normalized with that of the air/*z*-cut quartz. More details of the HD-SFG measurements can be found in the Supplementary Method of the Supplementary Information.

**qPlus-based AFM measurement.** All experiments were conducted using a homemade system that combines scanning tunneling microscopy (STM) and noncontact atomic force microscopy (nc-AFM). The qPlus sensor was equipped with a tungsten (W) tip, characterized by a spring constant of approximately 1800 N·m$^{-1}$, a resonance frequency of about 28.9 kHz, and a quality factor of around 60000. All AFM data were collected at 6 K under ultra-high vacuum conditions (<5×10$^{-10}$ Torr). High-resolution AFM images were acquired in constant-height mode. A carbon monoxide (CO) tip was prepared on an Au(111) surface and subsequently transferred to hBN surfaces. Initially, a bare W tip was positioned directly above a CO molecule on the Au(111) surface (100 mV, 10 pA). The current was then increased to 500 pA, enabling the CO molecule to transfer to the tip apex. The oscillation amplitude of the qPlus sensor ranged from 100 to 500 pm. Image processing was performed using Nanotec WSxM software. The drift in tip-sample distance was minimal, with fluctuations of less than 1 pm over 8 minutes, and the temperature stability of our system improved to 0.01 K over one hour. Fluctuations in amplitude and frequency shifts were limited to less than 4 pm and 30 mHz, respectively. These characteristics ensure stable, long-term high-resolution imaging.

**Machine learning-based molecular dynamics simulations.** All simulations were performed using a machine learning potential (MLP) based on the MACE architecture[63]. We use 128 invariant channels, a cutoff distance of 6 Å, and two message-passing layers, resulting in an effective receptive field of 12 Å. The final energy and force root-mean-square errors of the model developed were 0.6 meV/atom and 19.4 meV/Å, respectively. The MLP developed and validated (see Supplementary Methods S5) accurately represents the potential energy of the system and was trained using energies and forces obtained using the CP2K/Quickstep code[64]. We specifically used the revPBE-D3[65,66] functional as it accurately reproduces the structure and dynamics of liquid water[56,57,67] and its ionized products[68]. The Kohn-Sham orbitals of oxygen and hydrogen atoms are expanded using a TZV2P basis set, while those of boron and nitrogen are expanded using a DZVP basis set[69] (see Supplementary Methods S5), along with electronic structure settings consistent with previous work[55]. The final model was trained on 8,402 structures encompassing the diverse range of conditions sampled, ensuring robust accuracy across different system configurations (see Methods). All MD simulations were performed at a temperature of 300 K in the NVT ensemble with a time step of 0.5 fs (see Supplementary Methods S6). The systems (with no strain) were modeled using a 17.396 Å × 17.577 Å × 35.000 Å orthorhombic cell, containing 112 surface atoms, one $OH^-$ ion, and 169 water molecules under periodic boundary conditions. To prevent interactions between periodic images, a 15 Å vacuum was included in the $z$ direction, exceeding the model's receptive field. In total, 5.05 ns of free MD and 3.96 ns of constrained MD simulations were performed, ensuring robust and statistically converged results. Constrained MD simulations were carried out using LAMMPS package[70] and PLUMED[71], while free MD simulations were conducted using the ASE[72] software.

## Data availability

Source data are provided with this paper.

## Competing interests

The authors declare no competing financial interest.

## Author contributions

Y.W. and M.B. designed the study. H.L., A.E., and Y.W. prepared the samples. Y.W. and H.L. performed the HD-SFG measurements and data analysis. Z.Z, D.W., and Y.J. characterized the sample using qPlus-based AFM. X.R.A., K.F., C.S., and A.M. designed and analyzed the simulation data. X.R.A. conducted the simulations. T.T.

provided the high-quality hBN crystal. Y.W., M.B., X.R.A., and A.M. wrote the manuscript. All authors contributed to interpreting the results and refining the manuscript.

## Acknowledgments

We are grateful for the financial support from the MaxWater Initiative of the Max Planck Society. Funded by the European Union (ERC, n-AQUA, 101071937). Views and opinions expressed are however those of the author(s) only and do not necessarily reflect those of the European Union or the European Research Council Executive Agency. Neither the European Union nor the granting authority can be held responsible for them. K.D.F. acknowledges support from Schmidt Science Fellows, in partnership with the Rhodes Trust, and Trinity College, Cambridge. C.S. acknowledges financial support from the Deutsche Forschungsgemeinschaft (DFG, German Research Foundation) project number 500244608, as well as from the Royal Society grant number RGS/R2/242614. This work used the ARCHER2 UK National Supercomputing Service via the UK's HEC Materials Chemistry Consortium, funded by EPSRC (EP/F067496). We also utilized resources from the Cambridge Service for Data Driven Discovery (CSD3), supported by EPSRC (EP/T022159/1) and DiRAC funding, with additional access through a University of Cambridge EPSRC Core Equipment Award (EP/X034712/1). The Cirrus UK National Tier-2 HPC Service at EPCC, funded by the University of Edinburgh and the EPSRC (EP/P020267/1), also provided computational support.

# References

1. Robin, P. *et al.* Long-term memory and synapse-like dynamics in two-dimensional nanofluidic channels. *Science* **379**, 161–167 (2023).

2. Wang, Z. *et al.* Graphene oxide nanofiltration membranes for desalination under realistic conditions. *Nat. Sustain.* **4**, 402–408 (2021).

3. Li, Z. *et al.* Breakdown of the Nernst–Einstein relation in carbon nanotube porins. *Nat. Nanotechnol.* **8**, 177–183 (2023).

4. Secchi, E. *et al.* Massive radius-dependent flow slippage in carbon nanotubes. *Nature* **537**, 210–213 (2016).

5. Mouterde, T. *et al.* Molecular streaming and its voltage control in ångström-scale channels. *Nature* **567**, 87–90 (2019).

6. Keerthi, A. *et al.* Water friction in nanofluidic channels made from two-dimensional crystals. *Nat. Commun.* **12**, 3092 (2021).

7. Radha, B. *et al.* Molecular transport through capillaries made with atomic-scale precision. *Nature* **538**, 222–225 (2016).

8. Li, Z. *et al.* Ion transport and ultra-efficient osmotic power generation in boron nitride nanotube porins. *Sci. Adv.* **10**, eado8081 (2024).

9. Kavokine, N., Bocquet, M.-L. & Bocquet, L. Fluctuation-induced quantum friction in nanoscale water flows. *Nature* **602**, 84–90 (2022).

10. Wu, D. *et al.* Probing structural superlubricity of two-dimensional water transport with atomic resolution. *Science* **384**, 1254–1259 (2024).

11. Bocquet, L. & Barrat, J.-L. Flow boundary conditions from nano- to micro-scales. *Soft Matter* **3**, 685–693 (2007).

12. Thiemann, F. L., Schran, C., Rowe, P., Müller, E. A. & Michaelides, A. Water flow in single-wall nanotubes: oxygen makes it slip, hydrogen makes it stick. *ACS Nano* **16**, 10775–10782 (2022).

13. Tocci, G., Joly, L. & Michaelides, A. Friction of water on graphene and hexagonal boron nitride from ab initio methods: very different slippage despite very similar interface structures. *Nano Lett.* **14**, 6872–6877 (2014).

14. Ohto, T., Tada, H. & Nagata, Y. Structure and dynamics of water at water–graphene and water–hexagonal boron-nitride sheet interfaces revealed by ab initio sum-frequency generation spectroscopy. *Phys. Chem. Chem. Phys.* **20**, 12979–12985 (2018).

15. Das, B., Ruiz-Barragan, S. & Marx, D. Deciphering the properties of nanoconfined aqueous solutions by vibrational sum frequency generation spectroscopy. *J. Phys. Chem. Lett.* **14**, 1208–1213 (2023).

16. Hosseini, A., Masoud Yarahmadi, A., Azizi, S., Korayem, A. H. & Savary, R. Water molecules in boron nitride interlayer space: ice and hydrolysis in super confinement. *Phys. Chem. Chem. Phys.*

**26**, 21841–21849 (2024).

17. Siria, A. *et al.* Giant osmotic energy conversion measured in a single transmembrane boron nitride nanotube. *Nature* **494**, 455–458 (2013).

18. Cetindag, S. *et al.* Anomalous diffusive and electric-field-driven ion transport in boron-nitride nanotubes. Preprint at https://doi.org/10.26434/chemrxiv-2023-w82kh (2023).

19. Wang, R. *et al.* In-plane dielectric constant and conductivity of confined water. Preprint at https://doi.org/10.48550/arXiv.2407.21538 (2024).

20. Seal, A. & Govind Rajan, A. Modulating water slip using atomic-scale defects: friction on realistic hexagonal boron nitride surfaces. *Nano Lett.* **21**, 8008–8016 (2021).

21. Verma, A. K. & Sharma, B. B. Unveiling the impact of atomic-scale defects and surface roughness on interfacial properties in hexagonal boron nitride. *Surf. Interface.* **51**, 104646 (2024).

22. Yang, F. *et al.* Understanding the intrinsic water wettability of hexagonal boron nitride. *Langmuir* **40**, 6445–6452 (2024).

23. Grosjean, B. *et al.* Chemisorption of hydroxide on 2d materials from dft calculations: graphene versus hexagonal boron nitride. *J. Phys. Chem. Lett.* **7**, 4695–4700 (2016).

24. Grosjean, B., Bocquet, M.-L. & Vuilleumier, R. Versatile electrification of two-dimensional nanomaterials in water. *Nat. Commun.* **10**, 1656 (2019).

25. Comtet, J. *et al.* Direct observation of water-mediated single-proton transport between hBN surface defects. *Nat. Nanotechnol.* **15**, 598–604 (2020).

26. Stein, D., Kruithof, M. & Dekker, C. Surface-charge-governed ion transport in nanofluidic channels. *Phys. Rev. Lett.* **93**, 035901 (2004).

27. Wang, Y. *et al.* Heterodyne-detected sum-frequency generation vibrational spectroscopy reveals aqueous molecular structure at the suspended graphene/water interface. *Angew. Chem. Int. Ed.* **63**, e202319503 (2024).

28. Joly, L., Ybert, C., Trizac, E. & Bocquet, L. Liquid friction on charged surfaces: From hydrodynamic slippage to electrokinetics. *J. Chem. Phys.* **125**, 204716 (2006).

29. Du, Q., Superfine, R., Freysz, E. & Shen, Y. R. Vibrational spectroscopy of water at the vapor/water interface. *Phys. Rev. Lett.* **70**, 2313 (1993).

30. Du, Q., Freysz, E. & Shen, Y. R. Surface vibrational spectroscopic studies of hydrogen bonding and hydrophobicity. *Science* **264**, 826–828 (1994).

31. Shen, Y. R. & Ostroverkhov, V. Sum-frequency vibrational spectroscopy on water interfaces: polar orientation of water molecules at interfaces. *Chem. Rev.* **106**, 1140–1154 (2006).

32. Bonn, M., Nagata, Y. & Backus, E. H. G. Molecular structure and dynamics of water at the water–air interface studied with surface-specific vibrational spectroscopy. *Angew. Chem. Int. Ed.* **54**, 5560–5576 (2015).

33. Ji, N., Ostroverkhov, V., Tian, C. S. & Shen, Y. R. Characterization of vibrational resonances of

water-vapor interfaces by phase-sensitive sum-frequency spectroscopy. *Phys. Rev. Lett.* **100**, 096102 (2008).

34. Yamaguchi, S. & Tahara, T. Heterodyne-detected electronic sum frequency generation: "Up" versus "down" alignment of interfacial molecules. *J. Chem. Phys.* **129**, 101102 (2008).

35. Seki, T. *et al.* Real-time study of on-water chemistry: Surfactant monolayer-assisted growth of a crystalline quasi-2D polymer. *Chem* **7**, 2758–2770 (2021).

36. Wen, Y.-C. *et al.* Unveiling microscopic structures of charged water interfaces by surface-specific vibrational spectroscopy. *Phys. Rev. Lett.* **116**, 016101 (2016).

37. Nihonyanagi, S. (二本柳聡史), Yamaguchi, S. (山口祥一) & Tahara, T. (田原太平). Direct evidence for orientational flip-flop of water molecules at charged interfaces: A heterodyne-detected vibrational sum frequency generation study. *J. Chem. Phys.* **130**, 204704 (2009).

38. Ohno, K., Okimura, M., Akai, N. & Katsumoto, Y. The effect of cooperative hydrogen bonding on the OH stretching-band shift for water clusters studied by matrix-isolation infrared spectroscopy and density functional theory. *Phys. Chem. Chem. Phys.* **7**, 3005–3014 (2005).

39. Adhikari, A. Accurate determination of complex $\chi^{(2)}$ spectrum of the air/water interface. *J. Chem. Phys.* **143**, 124707 (2015).

40. Brown, M. G., Walker, D. S., Raymond, E. A. & Richmond, G. L. Vibrational sum-frequency spectroscopy of alkane/water interfaces: experiment and theoretical simulation. *J. Phys. Chem. B* **107**, 237–244 (2003).

41. Moore, F. G. & Richmond, G. L. Integration or segregation: how do molecules behave at oil/water interfaces? *Acc. Chem. Res.* **41**, 739–748 (2008).

42. Strazdaite, S., Versluis, J., Backus, E. H. G. & Bakker, H. J. Enhanced ordering of water at hydrophobic surfaces. *J. Chem. Phys.* **140**, 054711 (2014).

43. Zhang, Y., de Aguiar, H. B., Hynes, J. T. & Laage, D. Water structure, dynamics, and sum-frequency generation spectra at electrified graphene interfaces. *J. Phys. Chem. Lett.* **11**, 624–631 (2020).

44. Wang, Y. *et al.* Interfaces govern structure of angstrom-scale confined water. Preprint at https://doi.org/10.48550/arXiv.2310.10354 (2023).

45. Xu, Y., Ma, Y.-B., Gu, F., Yang, S.-S. & Tian, C.-S. Structure evolution at the gate-tunable suspended graphene–water interface. *Nature* **621**, 506–510 (2023).

46. Ward, R. N., Duffy, D. C., Davies, P. B. & Bain, C. D. Sum-frequency spectroscopy of surfactants adsorbed at a flat hydrophobic surface. *J. Phys. Chem.* **98**, 8536–8542 (1994).

47. Richmond, G. L. Molecular bonding and interactions at aqueous surfaces as probed by vibrational sum frequency spectroscopy. *Chem. Rev.* **102**, 2693–2724 (2002).

48. Wang, Y. *et al.* Direct probe of electrochemical pseudocapacitive ph jump at a graphene electrode**. *Angew. Chem. Int. Ed.* **62**, e202216604 (2023).

49. Ohno, P. E., Wang, H. & Geiger, F. M. Second-order spectral lineshapes from charged interfaces.

*Nat. Commun.* **8**, 1032 (2017).

50. Reddy, S. K. *et al.* Bulk contributions modulate the sum-frequency generation spectra of water on model sea-spray aerosols. *Chem* **4**, 1629–1644 (2018).

51. Litman, Y., Chiang, K.-Y., Seki, T., Nagata, Y. & Bonn, M. Surface stratification determines the interfacial water structure of simple electrolyte solutions. *Nat. Chem.* **16**, 644–650 (2024).

52. Wang, Y. *et al.* Chemistry governs water organization at a graphene electrode. *Nature* **615**, E1–E2 (2023).

53. Braunschweig, B., Eissner, S. & Daum, W. Molecular structure of a mineral/water interface: effects of surface nanoroughness of α-al2o3 (0001). *J. Phys. Chem. C* **112**, 1751–1754 (2008).

54. Dalstein, L., Potapova, E. & Tyrode, E. The elusive silica/water interface: isolated silanols under water as revealed by vibrational sum frequency spectroscopy. *Phys. Chem. Chem. Phys.* **19**, 10343–10349 (2017).

55. Advincula, X. R., Fong, K. D., Michaelides, A. & Schran, C. Protons accumulate at the graphene-water interface. Preprint at https://doi.org/10.48550/arXiv.2408.04487 (2025).

56. Gillan, M. J., Alfè, D. & Michaelides, A. Perspective: How good is DFT for water? *J. Chem. Phys.* **144**, 130901 (2016).

57. Marsalek, O. & Markland, T. E. Quantum dynamics and spectroscopy of ab initio liquid water: the interplay of nuclear and electronic quantum effects. *J. Phys. Chem. Lett.* **8**, 1545–1551 (2017).

58. Das, S., Bonn, M. & Backus, E. H. G. The surface activity of the hydrated proton is substantially higher than that of the hydroxide ion. *Angew. Chem. Int. Ed.* **58**, 15636–15639 (2019).

59. Das, S. *et al.* Nature of excess hydrated proton at the water–air interface. *J. Am. Chem. Soc.* **142**, 945–952 (2020).

60. McCaffrey, D. L. *et al.* Mechanism of ion adsorption to aqueous interfaces: Graphene/water vs. air/water. *Proc. Natl. Acad. Sci. U.S.A.* **114**, 13369–13373 (2017).

61. Moloney, E. G., Azam, Md. S., Cai, C. & Hore, D. K. Vibrational sum frequency spectroscopy of thin film interfaces. *Biointerphases* **17**, 051202 (2022).

62. Yang, S. *et al.* Nature of the electrical double layer on suspended graphene electrodes. *J. Am. Chem. Soc.* **144**, 13327–13333 (2022).

63. Batatia, I., Kovacs, D. P., Simm, G. N. C., Ortner, C. & Csanyi, G. MACE: Higher order equivariant message passing neural networks for fast and accurate force fields. in (2022).

64. Kühne, T. D. *et al.* CP2K: An electronic structure and molecular dynamics software package - Quickstep: Efficient and accurate electronic structure calculations. *J. Chem. Phys.* **152**, 194103 (2020).

65. Perdew, J. P., Burke, K. & Ernzerhof, M. Generalized gradient approximation made simple. *Phys. Rev. Lett.* **77**, 3865–3868 (1996).

66. Grimme, S., Antony, J., Ehrlich, S. & Krieg, H. A consistent and accurate ab initio parametrization

of density functional dispersion correction (DFT-D) for the 94 elements H-Pu. *J. Chem. Phys.* **132**, 154104 (2010).

67. Morawietz, T., Singraber, A., Dellago, C. & Behler, J. How van der Waals interactions determine the unique properties of water. *Proc. Natl. Acad. Sci. U.S.A.* **113**, 8368–8373 (2016).

68. Atsango, A. O., Morawietz, T., Marsalek, O. & Markland, T. E. Developing machine-learned potentials to simultaneously capture the dynamics of excess protons and hydroxide ions in classical and path integral simulations. *J. Chem. Phys.* **159**, 074101 (2023).

69. VandeVondele, J. & Hutter, J. Gaussian basis sets for accurate calculations on molecular systems in gas and condensed phases. *J. Chem. Phys.* **127**, 114105 (2007).

70. Thompson, A. P. *et al.* LAMMPS - a flexible simulation tool for particle-based materials modeling at the atomic, meso, and continuum scales. *Comput. Phys. Commun.* **271**, 108171 (2022).

71. Tribello, G. A., Bonomi, M., Branduardi, D., Camilloni, C. & Bussi, G. PLUMED 2: New feathers for an old bird. *Comput. Phys. Commun.* **185**, 604–613 (2014).

72. Hjorth Larsen, A. *et al.* The atomic simulation environment—a Python library for working with atoms. *J. Phys.: Condens. Matter* **29**, 273002 (2017).

# Supplementary information for:

# Spontaneous Surface Charging and Janus Nature of the Hexagonal Boron Nitride-Water Interface

Yongkang Wang[1*,#], Haojian Luo[1#], Xavier R. Advincula[2,3,4#], Zhengpu Zhao[5#], Ali Esfandiar[1,6], Da Wu[5], Kara D. Fong[2,4], Lei Gao[1], Arsh S. Hazrah[1], Takashi Taniguchi[7], Christoph Schran[3,4], Yuki Nagata[1], Lydéric Bocquet[6], Marie-Laure Bocquet[6], Ying Jiang[5], Angelos Michaelides[2,4], Mischa Bonn[1*]

[1] *Department of molecular spectroscopy, Max Planck Institute for Polymer Research, Ackermannweg 10, 55128 Mainz, Germany.*

[2] *Yusuf Hamied Department of Chemistry, University of Cambridge, Lensfield Road, Cambridge, CB2 1EW, UK.*

[3] *Cavendish Laboratory, Department of Physics, University of Cambridge, Cambridge, CB3 0HE, UK.*

[4] *Lennard-Jones Centre, University of Cambridge, Trinity Ln, Cambridge, CB2 1TN, UK.*

[5] *International Center for Quantum Materials, School of Physics, Peking University, Beijing 100871, China.*

[6] *Laboratoire de Physique de l'École Normale Supérieure, ENS, Université PSL, CNRS, Sorbonne Université, Université Paris Cité, F-75005 Paris, France.*

[7] *Research Center for Materials Nanoarchitectonics, National Institute for Materials Science, Tsukuba, Japan.*

[#]*Those authors contributed equally to this work.*

[*]*Correspondence to:* wangy3@mpip-mainz.mpg.de, bonn@mpip-mainz.mpg.de

**Supplementary Information contains:**

Supplementary Methods S1-S5

Supplementary Notes S1-S8

Figs. S1 to S14

Tables S1-S3

# Contents

## Supplementary Methods

### S1. Chemicals

All related chemicals of sodium chloride (NaCl), heavy water ($D_2O$), sodium hydroxide (NaOH), hydrochloride (HCl, 37%), concentrated sulfuric acid ($H_2SO_4$, 98%), 30 wt. % hydrogen peroxide solution ($H_2O_2$), ethanol, and acetone were purchased from Sigma-Aldrich and were all of analytical grade without further purification. Polydimethylsiloxane (PDMS) was provided by Dow, Inc. Deionized water was provided by a Milli-Q system (resistivity $\geq$ 18.2 M$\Omega$·cm and TOC $\leq$ 4 ppb). CVD-grown graphene on copper foils was purchased from Grolltex Inc. hBN crystals were obtained from International Center for Materials Nanoarchitectonics, National Institute for Materials Science 1-1 Namiki, Tsukuba 305-0044, Japan.

### S2. SiO$_2$ Substrate Preparation

Water-free $SiO_2$ substrates (10×10 ×1 mm$^3$, PI-KEM Ltd) were cleaned with acetone and 2-propanol sequentially. Prior to the transfer of hBN flakes, the $SiO_2$ substrate was subjected to an oxygen plasma treatment (300 W, 20 sccm $O_2$, and a duration of 10 minutes). This oxygen plasma treatment was to ensure surface cleanliness and enhance the adhesion with the hBN flakes during the transfer process.

### S3. hBN Sample Preparation

High-quality hBN flakes were exfoliated via mechanical cleavage using polydimethylsiloxane (PDMS) substrate (SYLGARD™ 184 Silicone Elastomer Kit, mixed at a 9:1 ratio of base to curing agent). To thin the hBN flakes, we repeatedly exfoliated the flakes using fresh PDMS substrates until the thickness was reduced to less than 100 nm. The final hBN flakes with larger than 200 × 200 μm² area were then identified using an optical microscope and dry-transferred onto an oxygen plasma-treated $SiO_2$ substrate. After flake preparation, a flat and clean region of approximately 150 × 150 μm² in size was identified and protected using an optical microscope and shadow mask. Then, a Cr/Au (3nm/100nm) film was deposited on the hBN crystals by electron gun evaporation to mark the identified area for the HD-SFG measurement and cover the edge of the flake and substrate regions. The preparation of the suspended graphene on the water surface was similar to Refs.[1,2] and was detailed in our recent work[3].

### S4. HD-SFG Measurement

HD-SFG measurements were performed on a non-collinear beam geometry with a Ti:Sapphire regenerative amplifier laser system. A detailed description can be found in Ref.[4,5]. The measurements were performed at the *ssp* polarization combination, where *ssp* denotes *s*-polarized SFG, *s*-polarized visible, and *p*-polarized IR beams. The power of the IR and visible beams was approximately 1 mW and 2 mW for measuring the hBN/water interface. We ensured that the IR and visible beam irradiations were far below the damage threshold of hBN (Supplementary Note S8). The IR, visible, and LO beams are directed at the sample (in $SiO_2$) at incidence angles of approximately 34°, 43°, and 41°, respectively. Each spectrum was acquired with an exposure time of 10 minutes and measured more than 6 times on average. All the HD-SFG spectra were measured in a dried air atmosphere to avoid spectral distortion due to water vapor. To obtain the phase information, the hBN/$H_2O$ HD-SFG signal at *ssp* polarization was normalized with the signal of hBN/$D_2O$ at *ssp* polarization at the same sample spot.

A description of our sample cell can be found in refs[4,5], and will not be elaborated here. For the ion concentration- and pH-dependent HD-SFG measurements, the flow cell was connected to a syringe pump for the supply of solutions. For each measurement, the cell was pumped with the salt solution for ~10 minutes and the processes were repeated three times before the HD-SFG measurement to avoid the memory effect.

**S5. Machine Learning Potential**

**Model Development.** To develop the MLP, we utilized training data from previous studies[6,7], which included datasets for the hBN-water interface and bulk water under various conditions. In addition, we incorporated configurations specifically targeting the chemisorbed and physisorbed states, as well as sampling the $OH^-$ ion in water layers farther from the hBN interface. To avoid the need for a homogeneous background charge, we included an $H_3O^+$ ion positioned far from its counterion to maintain charge neutrality—an approach previously shown to be effective[8,9]. This method eliminates dependencies on the simulation box volume, ensuring the robustness of our MLP. Moreover, it enables the model to accurately describe the behavior of the $H_3O^+$ ion. To further refine the MLP, we expanded our dataset to include pure water and neutral water containing protonic defects across various environments, including the air-water interface, the hBN-water interface with varying water layer thicknesses, and water confined between hBN sheets. Finally, we performed an additional round of active learning to optimize the model under these diverse conditions.

**Electronic Structure Settings.** The MLP was developed (and validated) using the energies and forces from the training data obtained at the DFT level. For this, we used the CP2K/Quickstep code[10]. We specifically used the revPBE-D3[11,12] functional as it accurately reproduces the structure and dynamics of liquid water[13–15] and its ionized products[8]. Atomic cores were represented using dual-space GTH pseudopotentials[16]. The Kohn-Sham orbitals of oxygen and hydrogen atoms were expanded using the TZV2P basis set, while the DZVP basis set was used for boron and nitrogen atoms[17]. Additionally, an auxiliary plane-wave basis with a cutoff of 1050 Ry was employed to represent the electron density.

**Model Validation.** We validated the model's ability to reproduce the reference method by quantifying the root-mean-square error (RMSE) of energies and forces using structures obtained from 250 ps MLP-based MD simulations. These simulations targeted the chemisorbed and physisorbed states.

To evaluate the accuracy and consistency of our results, we performed additional single-point DFT calculations on 800 randomly selected snapshots extracted from the MD simulations, computing their energies and forces. This conforms the additional set of test data to evaluate the MLP developed in this work.

To reduce computational costs involved in these DFT single-point calculations, we scaled down the system dimensions, setting the hBN lattice parameters to 13.047 Å × 12.555 Å as opposed to 17.396 Å × 17.577 Å. Unlike the MLP simulations, these configurations included only the $OH^-$ ion without a counterion, necessitating the application of a homogeneous background charge to maintain charge neutrality. To account for the energy shift introduced by this charge when comparing the energies to the MLP, we subtracted a constant energy offset from the MLP-predicted energies. Notably, this validation approach is particularly robust, as it directly assesses structures sampled from the MLP's potential energy surface, ensuring an accurate comparison with the reference method. As shown in Fig. S1 and Fig. S2, the MLP exhibits strong agreement with the reference DFT calculations for both the chemisorbed and physisorbed states, demonstrating its ability to accurately reproduce the underlying level of theory.

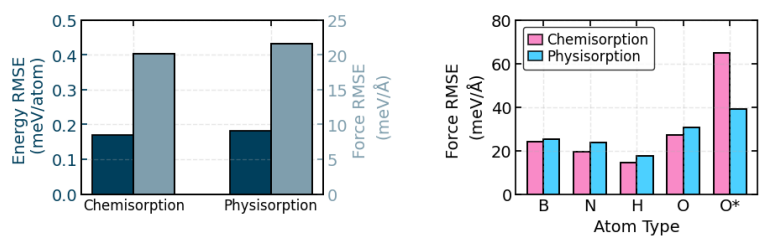

**Fig. S1 | RMSE of energies and forces predicted by the MLP compared to reference DFT calculations.** The force RMSE is further broken down by atom type, where $O^*$ denotes the oxygen atom in the $OH^-$ ion.

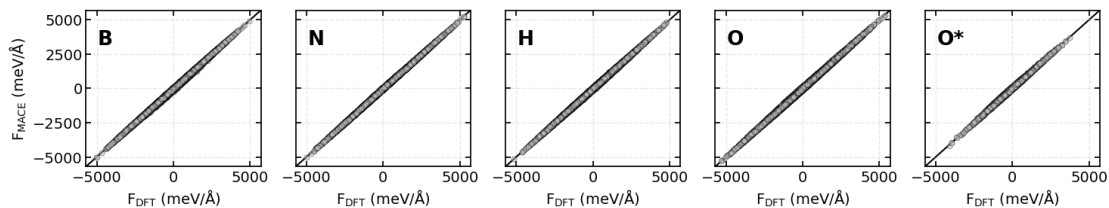

**Fig. S2 | Parity plots for the forces obtained using the MLP compared to reference DFT calculations broken down by atom type, where O* denotes the oxygen atom in the OH⁻ ion.**

### S1. Molecular Dynamics Simulations

All MD simulations were performed using the MLP at a temperature of 300 K under the NVT ensemble, with a time step of 0.5 fs. Simulations were conducted in orthorhombic cells with periodic boundary conditions applied in all three directions. The systems with no strain were modeled using a 17.396 Å × 17.577 Å × 35.000 Å orthorhombic cell, containing 112 surface atoms, one OH⁻ ion, and 169 water molecules under periodic boundary conditions. For the strained systems, we applied a ±2% variation along the $x$-axis, resulting in cell dimensions ranging from 17.048 Å × 17.577 Å × 35.000 Å to 17.744 Å × 17.577 Å × 35.000 Å. A representative snapshot of the simulated systems is shown in Fig. S3.

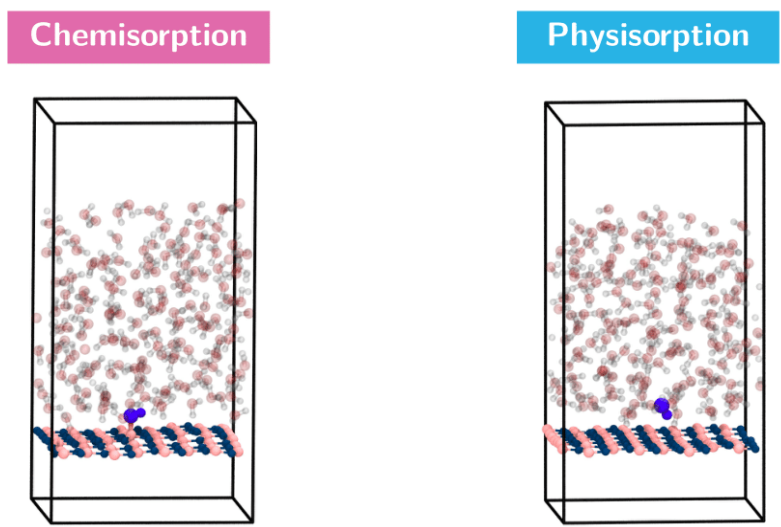

**Fig. S3 | Representative snapshot illustrating the dimensions of the systems studied.**

For the free MD simulations, thermalization was achieved using a Langevin thermostat with a friction coefficient of 2.5 ps⁻¹. Each simulation included a 50 ps equilibration phase followed by its corresponding production run.

For the constrained MD simulations used in restrained umbrella sampling, thermalization was achieved using a Nosé-Hoover thermostat with a damping constant of 0.05 ps. A total of 33 umbrella windows were sampled, each undergoing a 45 ps equilibration phase followed by a 75 ps production run. The reported PMF profile was obtained using umbrella integration.

In each umbrella window, the oxygen atom of the OH⁻ ion (O*) was restrained at different target heights above a fixed B atom, while the rest of the hBN interface remained fully flexible. The restraining potential applied took the form:

$$U_{bias,1}(z) = \frac{k_{bias,1}}{2}(z - z_0)^2, \tag{S1}$$

where $z$ is the instantaneous height of the $O^*$ above the hBN sheet, defined as the distance between O and the fixed B atom. The force constant is set to $k_{bias,1} = 150$ kcal/mol/Å. To avoid proton hopping, we restrained the hydrogen coordination value of the O* around a target value $n_0$ (here, this is 1.0) using a harmonic potential of the form,

$$U_{bias,2}(z) = \frac{k_{bias,2}}{2}(n_{O*-H} - n_0)^2, \tag{S2}$$

where $k_{bias,2} = 400$ kcal/mol per coordination unit squared and

$$n_{O*-H} = \sum_{i=1}^{N} \frac{1 - \left(\frac{r_i}{R_0}\right)^{12}}{1 - \left(\frac{r_i}{R_0}\right)^{20}}, \tag{S3}$$

where $i$ iterates over all the hydrogens in the simulation box, $r_i$ is the distance between a hydrogen $i$ and O*, and $R_0$ is a switching distance (1.2 Å).

# Supplementary Note

## S1. Cleanness of the Prepared hBN Surface

To confirm the hBN surface is clean and flat within the HD-SFG probe region (the diameter of the laser spot is around 100 μm), we conducted AFM measurements on the hBN surface across a 100×100 μm² region. The large-area surface morphology of the hBN samples was measured using an atomic force microscope (AFM, Bruker, JPK) working in the noncontact mode. We used a silicon cantilever (OPUS-240AC, f = 70 kHz, k = 2 Nm$^{-1}$) for the measurement. The AFM data shows that the hBN surface appears clean and atomically flat with no visible layered step edges within the SFG probed region, showing an RMS surface roughness ($R_q$) measuring around 0.7 Å (Fig. S4).

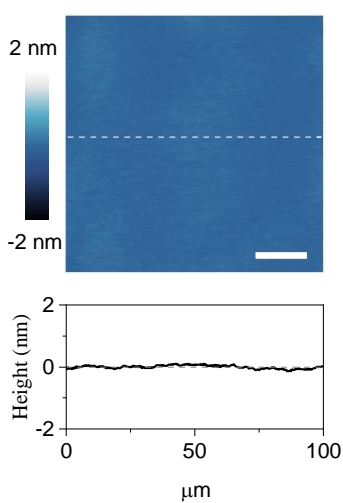

**Fig. S4 | Characterization of the hBN surface.** AFM height image of the hBN surface. The scale bar is 20 μm. The bottom panel shows the height profiles along the white dashed lines in the AFM height image. The dashed grey lines in the height profiles indicate zero lines. $R_q$ values were calculated across the whole scan area.

## S2. Screening of Substrate Effect

The supporting substrate may influence interfacial water arrangement at the substrate-supported two-dimensional materials/water interface, such as substrate-supported monolayer graphene[4,5,18–20]. To avoid the substrate effect, we prepared approximately 100 nm thick hBN flakes. To confirm that the substrate effect is efficiently screened by the approximately 100 nm thick hBN flake, we prepared the hBN flakes using different substrates (SiO$_2$ and CaF$_2$) with different polarities and measured the $\text{Im}(\chi^{(2)}_{\text{BN}})$ spectra. The data shown in Fig. S5 confirms that the substrate effect is effectively screened by the approximately 100 nm thick hBN flake.

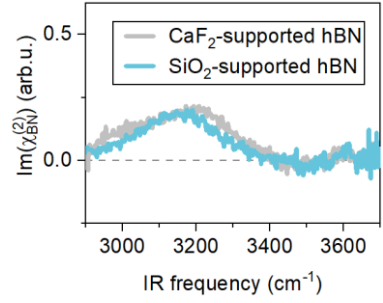

**Fig. S5 | Screening of substrate effect.** Experimental $\text{Im}\left(\chi_{\text{BN}}^{(2)}\right)$ spectra of pure water obtained at CaF$_2$- and SiO$_2$-supported hBN/water interfaces. The grey dashed line represents a zero line.

**S3. Phase Measurement and Fresnel Factor**

**Phase Measurement**. To obtain the phase information, the HD-SFG signal ($\chi_{\text{eff},ssp,\text{SiO}_2-\text{hBN}/\text{H}_2\text{O}}^{(2)}$) of the hBN/H$_2$O interface at *ssp* polarization was normalized by the signal ($\chi_{\text{eff},ssp,\text{SiO}_2-\text{hBN}/\text{D}_2\text{O}}^{(2)}$) of hBN/D$_2$O interface at *ssp* polarization. The measured SFG response ($\chi_{ssp,\text{measured}}^{(2)}$) is thus given by:

$$\chi_{ssp,\text{measured}}^{(2)} = \frac{\chi_{\text{eff},ssp,\text{SiO}_2-\text{hBN}/\text{H}_2\text{O}}^{(2)}}{\chi_{\text{eff},ssp,\text{SiO}_2-\text{hBN}/\text{D}_2\text{O}}^{(2)}}. \quad (S4)$$

**Fresnel Factor Correction**. While bulk hBN is SFG-inactive, its surface can exhibit significant non-resonant second-order susceptibility ($\chi_{yyz,\text{hBN}}^{(2)}$), primarily arising from the outermost hBN layer where inversion symmetry is broken. This response is purely real[21–23]. Although both hBN and D$_2$O response are pure real, for the thin film interface, the Fresnel factor may influence both the amplitude and phase of $\chi_{ssp,\text{measured}}^{(2)}$. To account for these effects, a Fresnel factor correction was conducted. The SiO$_2$-supported hBN/water interface consists of three bulk media, represented by refractive indices $n_1$, $n_2$, and $n_3$, and two interfacial regions, represented by refractive indices $n'$ and $n''$, as illustrated in Fig. S6. In such a three-phase system, the effective SFG response ($\chi_{\text{eff},ssp}^{(2)}$) is expressed as[24]:

$$\chi_{\text{eff},ssp}^{(2)} = F^{12}\left(\chi_{yyz,\text{hBN}}^{(2)} + \chi_{yyz,\text{EQ}}^{(2)}\right) + F^{23}\left(-\chi_{yyz,\text{hBN}}^{(2)} - \chi_{yyz,\text{EQ}}^{(2)}\right) + F^{23}\chi_{yyz}^{(2)}, \quad (S5)$$

where $F^{12}$, $F^{23}$ are Fresnel factors for the SiO$_2$/hBN ($z = 0$) and hBN/water ($z = d$) interfaces respectively. $\chi_{yyz,\text{EQ}}^{(2)}$ accounts for the non-resonant response at the two interfaces, primarily originating from the electric quadrupole contribution[25,26] which is purely real with its amplitude not highly sensitive to the refractive indices of the two bulk media forming the interface. We assumed that the amplitudes of $\chi^{(2)}_{yyz,\text{hBN}}$ and $\chi^{(2)}_{yyz,\text{EQ}}$ remain the same for the SiO$_2$/hBN ($z=0$) and hBN/water ($z=d$) interfaces, but with opposite sign.

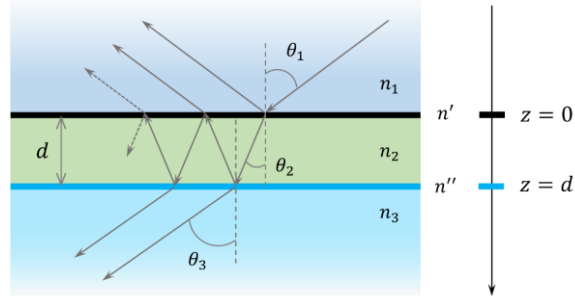

**Fig. S6 | Schematic diagram illustrating multiple reflections of an incident light in the hBN film.**

Since the non-resonant background ($\chi^{(2)}_{yyz,\text{NR}} = \chi^{(2)}_{yyz,\text{hBN}} + \chi^{(2)}_{yyz,\text{EQ}}$) is explicitly included in Eq. S5, $\chi^{(2)}_{yyz}$ represents the *yyz* component of the resonant SFG signal for H$_2$O, which is absent for D$_2$O. As such, from Eq. S4 and S5, the $\chi^{(2)}_{yyz}$ signal at the SiO$_2$-supported hBN/water interface can be expressed as:

$$\chi^{(2)}_{yyz} = \left(\chi^{(2)}_{ssp,\text{measured}} - \frac{F^{12}_{\text{H}_2\text{O}} - F^{23}_{\text{H}_2\text{O}}}{F^{12}_{\text{D}_2\text{O}} - F^{23}_{\text{D}_2\text{O}}}\right)\frac{(F^{12}_{\text{D}_2\text{O}} - F^{23}_{\text{D}_2\text{O}})\chi^{(2)}_{yyz,\text{NR}}}{F^{23}_{\text{H}_2\text{O}}}. \tag{S6}$$

The Fresnel factors $F^{12}$ and $F^{23}$ were calculated via[24]:

$$F^{12} = \left(1 + \tilde{r}_{s,12}(\omega_{\text{SF}})\right) \times \left(1 + \tilde{r}_{s,12}(\omega_{\text{vis}})\right) \times \left(1 + \tilde{r}_{p,12}(\omega_{\text{IR}})\right)\left(\frac{n_1}{n'}\right)^2 \sin\theta_1, \tag{S7}$$

$$F^{23} = \tilde{t}_{s,23}(\omega_{\text{SF}}) \times \tilde{t}_{s,23}(\omega_{\text{vis}}) \times \tilde{t}_{p,23}(\omega_{\text{IR}})\frac{n_1 n_3}{(n'')^2}\sin\theta_1, \tag{S8}$$

where $\tilde{r}$ and $\tilde{t}$ denote the overall reflection and transmission coefficients at the SiO$_2$-supported hBN/water interface, incorporating multiple reflections within the thin hBN film[24]. They are calculated using Eq. S9 and S10[24]. $\omega_{\text{SF}}$, $\omega_{\text{vis}}$, and $\omega_{\text{IR}}$ denote the frequencies of the SF, visible, and IR light, respectively. $\theta$ denotes the incident angle. The subscripts *s* and *p* indicate the *s*-polarized and *p*-polarized light, respectively. The indices 1, 2, and 3 correspond to the respective media. Additionally, $n'$ and $n''$ represent the two interfacial dielectric constants. In this work, the Slab model is employed to describe the two interfacial dielectric constants ($n'$ and $n''$) which is a commonly used approach for complex multilayer systems under the ssp polarization combination[24].

$$\tilde{r} = r_{12} + \frac{t_{12}r_{23}t_{21}e^{i\Delta\phi}}{1 - r_{21}r_{23}e^{i\Delta\phi}}, \tag{S9}$$

$$\tilde{t} = \frac{t_{12}t_{23}e^{i\Delta\phi/2}}{1 - r_{23}r_{21}e^{i\Delta\phi}}, \tag{S10}$$

where $r$ and $t$ are the Fresnel reflection coefficient and Fresnel transmission coefficient at single interface. Notably, Eq. S9 is used to obtain $\tilde{r}_s$ and $\tilde{r}_p$ depending on whether $r_s$ and $t_s$ or $r_p$ and $t_p$ are utilized. This also applies to Eq. S10. $\Delta\phi = 4\pi d/\lambda\, n_2 \cos\theta_2$ is the propagation phase shift for light ($\lambda$, wavelength in a vacuum) passing through the thin hBN layer of thickness $d = 100$ nm. Notably, the thickness of the hBN flake was specifically chosen to ensure that the SFG primarily probes the hBN/water ($z = d$) interface, where $F^{23}$ dominates over $F^{12}$.

**Table S1. Refractive indexes used to calculate the Fresnel factors.**

| Refractive index $n$ | SF (~635 nm) | Vis (800 nm) | IR (3000 nm) |
|---|---|---|---|
| hBN | 2.13 | 2.10 | 2.00 |
| SiO$_2$ | 1.46 | 1.45 | 1.41 |
| D$_2$O | 1.33 | 1.33 | 1.25 |

To get $\chi^{(2)}_{yyz}$ via Eq. S6, knowledge of the non-resonant background ($\chi^{(2)}_{yyz,\text{NR}} = \chi^{(2)}_{yyz,\text{hBN}} + \chi^{(2)}_{yyz,\text{EQ}}$) is required. Previous studies have shown that $\chi^{(2)}_{yyz,\text{hBN}}$ is pure real[27], with its amplitude ranging from 0 to $1.5 \times 10^{-20}$ m$^2$/V, depending on the thickness and crystal orientation around the $z$-axis of the hBN[21–23]. In our SFG measurements, the hBN crystal orientation around the $z$-axis was manually optimized to be close to the maximum intensity of the homodyne SFG signal at the SiO$_2$-supported hBN/D$_2$O interface. This approach enhanced the hBN response, improving overall SFG signal stability and minimizing laser instability effects. Nevertheless, the exact amplitude of $\chi^{(2)}_{yyz,\text{hBN}}$ remains unknown, we instead inferred $\chi^{(2)\prime}_{yyz} = \chi^{(2)}_{yyz}/\chi^{(2)}_{yyz,\text{NR}}$ from Eq. S6. Using the parameters listed in Table S1 and Eq. S6-S10, the inferred $\chi^{(2)\prime}_{yyz}$ is shown in Fig. S7. Notably, for H$_2$O, frequency-dependent refractive index was employed[28,29].

The results confirm that the Fresnel factors do not alter the main conclusion: the hBN surface is negatively charged upon contacting water.

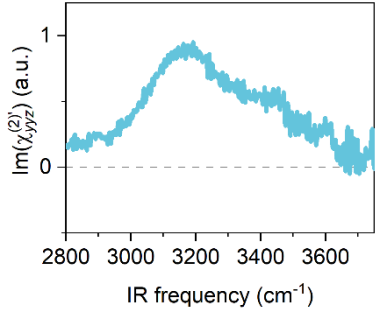

**Fig. S7 | Effect of Fresnel Factor.** Inferred $\text{Im}\left(\chi_{yyz}^{(2)\prime}\right)$ spectrum of pure water. The grey dashed line represents a zero line.

**S4. Effect of Carbonate from CO₂ Dissociation**

To examine the effect of carbonate due to $CO_2$ dissociation in water, we measured the $\text{Im}(\chi_{BN}^{(2)})$ spectrum using Ar-purged pure water. The data shown in Fig. S8 confirm that carbonate is not responsible for the negative charging of the hBN surface upon contacting water.

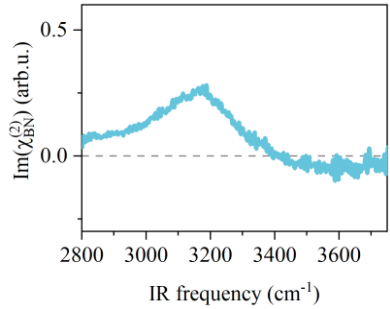

**Fig. S8 | Effect of carbonate.** Experimental $\text{Im}\left(\chi_{BN}^{(2)}\right)$ spectrum of Ar-purged pure water. The grey dashed line represents a zero line.

**S5. Determination of Surface Charge Density**

At charged interfaces, observed $\chi^{(2)}$ is given as the sum of the surface contribution ($\chi_s^{(2)}$-term) and the DC field-induced bulk contribution ($\chi^{(3)}$-term)[5,30–32]:

$$\chi^{(2)}(\sigma_0, c) = \chi_s^{(2)} + \chi^{(3)}\phi(\sigma_0, c)\frac{\kappa(c)}{\kappa(c) - i\Delta k_z}, \tag{S11}$$

where $\sigma_0$ is the net surface charge density at the charged interface, $\chi^{(3)}$ primarily represents the third-order nonlinear susceptibility originating from bulk water, $\phi$ is the electrostatic potential, $c$ is the ion strength, $\kappa$ is the inverse of Debye screening length, and $\Delta k_z \approx 1/25$ nm$^{-1}$ (@3300 cm$^{-1}$) is the phase-mismatch of the SF, visible, and IR beams in the depth direction. Assuming that the $\chi_s^{(2)}$-term is insensitive to the ion strength[30,33] and ions (⩽100 mM NaCl) do not affect the hBN surface charging, the $\chi^{(2)}$ spectral changes upon ion strength changes primarily arise from the variation of the $\chi^{(3)}$-term at the charged interface. This allows for determination of the $\chi^{(3)}$ spectrum from the differential spectrum $\Delta\chi^{(2)} = \chi^{(2)}(\sigma_0, c_1) - \chi^{(2)}(\sigma_0, c_2)$. To this end, we measured the $\chi_{BN}^{(2)}(\sigma_0, c)$ spectra at three different ion strengths at neutral pH~6. We used the ion strengths of $c_1 = 1\ \mu M$, $c_2 = 10\ mM$, and $c_3 = 100\ mM$ (see Fig. S9a). We then obtained $\sigma_0$ from the differential spectra $\Delta\chi_{BN}^{(2)}(\sigma_0, c_i, c_j) = \chi_{BN}^{(2)}(\sigma_0, c_i) - \chi_{BN}^{(2)}(\sigma_0, c_j)$ within the Gouy-Chapman model[34] via:

$$\frac{\Delta\chi^{(2)}(\sigma_0, c_1, c_3)}{\Delta\chi^{(2)}(\sigma_0, c_2, c_3)} = \frac{\left(\frac{\phi(\sigma_0, c_1)\kappa(c_1)}{\kappa(c_1) - i\Delta k_z} - \frac{\phi(\sigma_0, c_3)\kappa(c_3)}{\kappa(c_3) - i\Delta k_z}\right)}{\left(\frac{\phi(\sigma_0, c_2)\kappa(c_2)}{\kappa(c_2) - i\Delta k_z} - \frac{\phi(\sigma_0, c_3)\kappa(c_3)}{\kappa(c_3) - i\Delta k_z}\right)}. \tag{S12}$$

With known $\sigma_0$, the $\chi^{(3)}$ spectrum was obtained via:

$$\Delta\chi^{(2)}(\sigma_0, c_2, c_3) = \chi^{(3)}\left(\frac{\phi(\sigma_0, c_2)\kappa(c_2)}{(\kappa(c_2) - i\Delta k_z)} - \frac{\phi(\sigma_0, c_3)\kappa(c_3)}{(\kappa(c_3) - i\Delta k_z)}\right). \tag{S13}$$

The obtained $\chi^{(3)}$ spectrum is shown in Fig. S9b. The lineshape of the spectrum is consistent with that reported in Refs.[30]. Once $\chi^{(3)}$ is known, Eq. S13 allows us to estimate $\sigma_0$ at different pH values. The differential spectra $\text{Im}(\Delta\chi_{BN}^{(2)})$ at different pH values are presented in Fig. S9c and corresponding inferred $\sigma_0$ are shown in Fig. 3b.

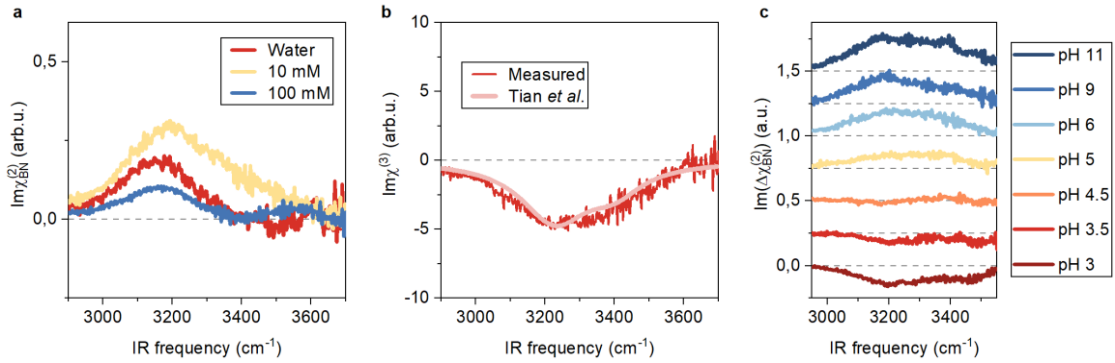

**Fig. S9 | Measurement of $\chi^{(3)}$ spectrum. a**. Experimental $\text{Im}(\chi_{BN}^{(2)})$ spectra at different ion strengths at pH~6. The dashed line represents the zero line. **b**. Comparison of the measured $\text{Im}(\chi^{(3)})$ spectrum and that reported in Ref.[30]. Note that $\text{Im}(\chi^{(3)})$ spectrum from Ref.[30] is rescaled in a way that the peak amplitude at ~3250 cm$^{-1}$ has the same value as that in our data. **c**.

Experimental Im$\left(\Delta\chi_{BN}^{(2)}\right)$ spectra at different pH values obtained from $\Delta\chi_{BN}^{(2)} = \chi_{BN}^{(2)}(\sigma_0, c_3 = 10\text{ mM}) - \chi_{BN}^{(2)}(\sigma_0, c_4 = 100\text{ mM})$. The grey dashed lines in (**a-c**) represent zero lines.

We note that the method used to estimate $\sigma_0$ relies on the Gouy-Chapman-Stern model[5,30–32], which assumes that the $\chi_s^{(2)}$-term is insensitive to ion strength and that ions (≤100 mM NaCl) do not influence the charging of the hBN surface. To examine the ion concentration effect, we also inferred $\sigma_0$ from $\Delta\chi_{BN}^{(2)}$ obtained from the SFG signal of 1 mM and 100 mM NaCl. The Im$\left(\Delta\chi_{BN}^{(2)}\right)$ data is shown in Fig. S10 and inferred $\sigma_0$ is -10 mC/m², slightly smaller than that inferred from difference spectrum $\Delta\chi_{BN}^{(2)}$ between 10 mM and 100 mM NaCl solutions. This analysis indicates that the NaCl ion concentration influences the surface charging of hBN; however, it does not alter our main conclusion that hBN undergoes spontaneous negative surface charging upon contact with water.

Accurately estimating $\sigma_0$ requires further refinement of the Gouy-Chapman-Stern model to account for both surface and bulk contributions in the SFG signal, such as ion-induced surface discharging[35,36], which remains a hot topic at the current stage[37]. This level of detail is beyond the scope of the present study and warrants additional investigation in future work.

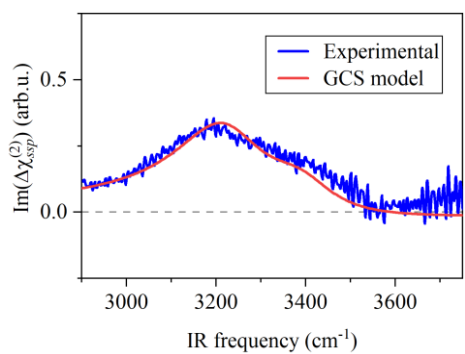

**Fig. S10 | Influence of ions on hBN surface charging.** Experimental difference spectrum Im$\left(\Delta\chi_{BN}^{(2)}\right)$ between 1 mM and 100 mM NaCl solutions, compared with a calculated spectrum based on the Gouy-Chapman-Stern theory for $\sigma_0 = -10$ mC/m². The grey dashed line represents a zero line.

## S6. qPlus-based AFM Data

To ensure the absence of defects on the hBN surface, we conducted qPlus-based AFM measurements over different randomly selected regions. Consistent with the data shown in Fig. 1f-h, the constant-height, high-resolution AFM images of the hBN surface from another randomly selected region reveal a clean surface with a perfect hexagonal honeycomb structure without any defects over an area of 100 nm² (Fig. S11).

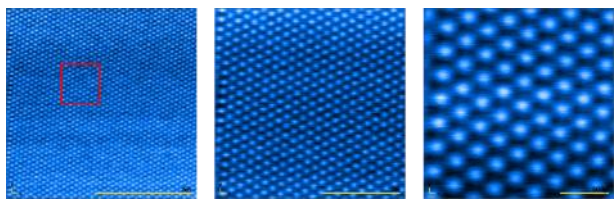

**Fig. S11 | qPlus-based AFM characterization of hBN. a**. Constant-height AFM image of the hBN surface. **b** and **c**. Zoomed-in AFM images from (**a**) with B and N atoms marked. The scale bars are 5 nm, 2 nm, and 0.5 nm, respectively.

**S7. Stability of the Chemisorbed and Physisorbed States**

To gain deeper insights into the stability of the chemisorbed and physisorbed states, we conducted additional free MD simulations, tracking the desorption time from the chemisorbed state (Table S2).

To further quantify the low free energy barrier between these states, we estimated the activation energy ($E_a$) using an Arrhenius-like expression:

$$E_a = k_B T \, ln(A\tau), \tag{S14}$$

where $E_a$ is the activation energy, $k_B$ is the Boltzmann constant, $T$ is the temperature, $A$ is the pre-exponential factor (assumed to be $10^{13}$ s$^{-1}$), and $\tau$ is the desorption time obtained from our simulations. This estimation provides a direct connection between the observed timescales and the energetic barriers governing the transition between adsorption states.

**Table S2. Free MD simulations starting from the chemisorbed state.**

| Run | Time for chemisorbed to physisorbed transition (ps) | Estimated activation energy, $E_a$ (eV) |
|---|---|---|
| #1 | 85 | 0.17 |
| #2 | 370 | 0.21 |
| #3 | 80 | 0.17 |
| #4 | 1685 | 0.25 |
| #5 | 750 | 0.23 |
| #6 | 460 | 0.22 |
| #7 | 1270 | 0.24 |

As shown in Table S2, the estimated barriers align well with the barriers obtained from the PMF in Fig. 2a, further validating our approach. Similarly, we can also provide an estimate of the time required by an OH⁻ ion to transition between the physisorbed and chemisorbed states. This can be done by rearranging Eq. S14 for $\tau$, where $E_a$ is obtained from the PMF in Fig. 2a. The data is shown in Table S3.

**Table S3. Estimated times required by an OH⁻ ion to transition between the states.**

| | $E_a$ | $\tau$ |
|---|---|---|
| Physisorbed to chemisorbed | 0.36 eV | 111.62 ns |
| Chemisorbed to physisorbed | 0.25 eV | 1.58 ns |

The comparable stability of the chemisorbed and physisorbed states suggests that nuclear quantum effects (NQEs) or exchange-correlation (XC) functional dependency may play a crucial role.

In the case of NQEs, their primary contribution can be approximated through the zero-point energy (ZPE), which, to a first-order approximation, is given by:

$$\text{ZPE} = \frac{1}{2}\hbar\omega, \tag{S15}$$

where $\omega$ represents the vibrational frequency. This frequency can be estimated using:

$$\omega = \sqrt{\frac{k}{m}}, \tag{S16}$$

where $k$ is the force constant, which can be determined by fitting a harmonic potential to the potential of mean force (PMF) in Fig. 2a, as done above. Substituting appropriately in Eq. S15, we calculate the ZPE for the chemisorbed state to be 0.0931 eV, while for the physisorbed state, it is 0.0275 eV. This difference highlights the steeper free energy well and higher vibrational frequency of the chemisorbed state compared to the physisorbed state. This is an admittedly crude semi-quantitative estimate of NQEs, neglecting for example the zero point energy of the other vibrational models in the system as well as any potential anharmonic effects. However, the magnitude of the ZPE difference between the two states implies that their relatively stability is unlikely to be greatly affected by NQEs.

The DFT XC functional is always an important consideration with simulating aqueous systems[13]. We have chosen the revPBE-D3 functional here as it accurately reproduces the structure and dynamics of liquid water[13–15] and its ionized products[8]. However, revPBE-D3 is a generalized gradient approximation (GGA) functional and it is well known that GGA functionals can overestimate electrostatic contributions to binding energies[38] and introduce delocalization errors[39]. These problems are largely ameliorated with hybrid functionals in which a fraction of exact (Hartree-Fock) exchange is introduced. To investigate this issue, we compared the system's total energies with the $OH^-$ ion in either the chemisorbed or the physisorbed state, using the revPBE-D3 and hybrid revPBE0-D3 functionals. For this analysis, we selected 300 configurations for the chemisorbed state and 300 configurations for the physisorbed state from MD simulations.

As shown in Fig. S12, both the revPBE-D3 and hybrid revPBE0-D3 functionals predict similar total energies between the physisorbed and chemisorbed states. In particular, revPBE0-D3 shifts the stability toward the chemisorbed state by 0.039 eV, making it more favorable. This effect is opposite to the influence of NQEs, which instead slightly stabilized the physisorbed state. As a result, these two contributions cancel each other to some extent, further highlighting the competitive balance between these states and reinforcing that these factors do not alter the main conclusions of our work.

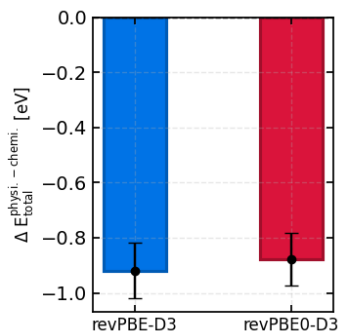

**Fig. S12 | Difference in the total energy of the system when the OH⁻ ion is in either the chemisorbed or the physisorbed state obtained with the revPBE-D3 functional (the functional the MLP used in the main text is trained on) and the hybrid revPBE0-D3 functional.** The total energy of the chemisorbed state serves as the reference (set to zero). Error bars represent the standard deviation from 300 sampled configurations for each state.

Lastly, to further explore the differences between the chemisorbed and physisorbed states, we examined the diffusive behavior of the OH⁻ ion in both configurations, as shown in Fig. S13. Our analysis reveals clear differences in their dynamics. In the chemisorbed state, OH⁻ remains largely fixed to the boron atom it is bonded to, exhibiting minimal movement. In contrast, in the physisorbed state, OH⁻ shows in-plane mobility, allowing it to diffuse more freely along the surface. This difference in mobility is particularly relevant for nanoscale friction on hBN.

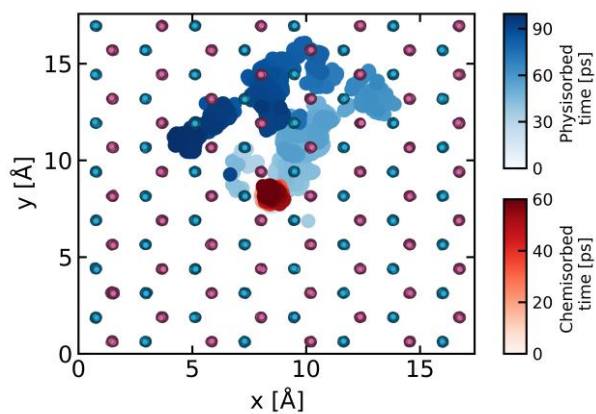

**Fig. S13 | In-plane motion of the OH⁻ ion in the chemisorbed and physisorbed states.** The positions are represented by the oxygen of the hydroxide ion. The color gradient represents the time in picoseconds for each state. In this simulation, the OH⁻ started in the chemisorbed state and transitioned to the physisorbed state at approx. 60 ps.

**S8. Fluence-independent SFG Signal**

To ensure the usage of 1 mW for IR (~3.3 μm) and 2 mW visible (800 nm) pulses do not damage the hBN sample, we compared the $\text{Im}(\chi_{BN}^{(2)})$ signals measured at different fluences. If the IR and visible pulses do damage the hBN sample, more defects (charges) are expected upon increasing the pulse power. The data displayed in Fig. S14 shows the water arrangement remains the same within the experimental uncertainty by increasing the power of IR and visible pulses, showing the IR and visible pulses do not introduce defects on the hBN surface.

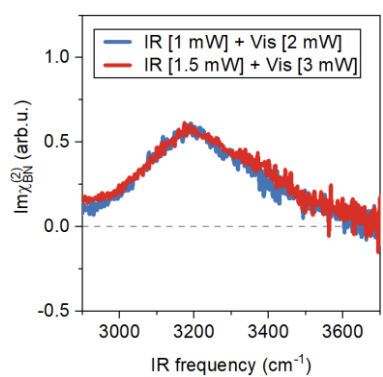

**Fig. S14 | Stability of hBN in contact with water under laser irradiation.** Experimental $\text{Im}(\chi_{BN}^{(2)})$ spectra obtained for water (10 mM NaCl) at pH~6 under different IR/vis fluences.

# References

1. Xu, Y., Ma, Y.-B., Gu, F., Yang, S.-S. & Tian, C.-S. Structure evolution at the gate-tunable suspended graphene–water interface. *Nature* **621**, 506–510 (2023).

2. Yang, S. *et al.* Nature of the electrical double layer on suspended graphene electrodes. *J. Am. Chem. Soc.* **144**, 13327–13333 (2022).

3. Wang, Y. *et al.* Heterodyne-detected sum-frequency generation vibrational spectroscopy reveals aqueous molecular structure at the suspended graphene/water interface. *Angew. Chem. Int. Ed.* **63**, e202319503 (2024).

4. Wang, Y. *et al.* Chemistry governs water organization at a graphene electrode. *Nature* **615**, E1–E2 (2023).

5. Wang, Y. *et al.* Direct probe of electrochemical pseudocapacitive ph jump at a graphene electrode**. *Angew. Chem. Int. Ed.* **62**, e202216604 (2023).

6. Thiemann, F. L., Schran, C., Rowe, P., Müller, E. A. & Michaelides, A. Water flow in single-wall nanotubes: oxygen makes it slip, hydrogen makes it stick. *ACS Nano* **16**, 10775–10782 (2022).

7. Ravindra, P., Advincula, X. R., Schran, C., Michaelides, A. & Kapil, V. Quasi-one-dimensional hydrogen bonding in nanoconfined ice. *Nat. Commun.* **15**, 7301 (2024).

8. Atsango, A. O., Morawietz, T., Marsalek, O. & Markland, T. E. Developing machine-learned potentials to simultaneously capture the dynamics of excess protons and hydroxide ions in classical and path integral simulations. *J. Chem. Phys.* **159**, 074101 (2023).

9. Advincula, X. R., Fong, K. D., Michaelides, A. & Schran, C. Protons accumulate at the graphene-water interface. Preprint at https://doi.org/10.48550/arXiv.2408.04487 (2025).

10. Kühne, T. D. *et al.* CP2K: An electronic structure and molecular dynamics software package - Quickstep: Efficient and accurate electronic structure calculations. *J. Chem. Phys.* **152**, 194103 (2020).

11. Perdew, J. P., Burke, K. & Ernzerhof, M. Generalized gradient approximation made simple. *Phys. Rev. Lett.* **77**, 3865–3868 (1996).

12. Grimme, S., Antony, J., Ehrlich, S. & Krieg, H. A consistent and accurate ab initio parametrization of density functional dispersion correction (DFT-D) for the 94 elements H-Pu. *J. Chem. Phys.* **132**, 154104 (2010).

13. Gillan, M. J., Alfè, D. & Michaelides, A. Perspective: How good is DFT for water? *J. Chem. Phys.* **144**, 130901 (2016).

14. Morawietz, T., Singraber, A., Dellago, C. & Behler, J. How van der Waals interactions determine the unique properties of water. *Proc. Natl. Acad. Sci. U.S.A.* **113**, 8368–8373 (2016).

15. Marsalek, O. & Markland, T. E. Quantum dynamics and spectroscopy of ab initio liquid water: the interplay of nuclear and electronic quantum effects. *J. Phys. Chem. Lett.* **8**, 1545–1551 (2017).

16. Goedecker, S., Teter, M. & Hutter, J. Separable dual-space Gaussian pseudopotentials. *Phys. Rev. B* **54**, 1703–1710 (1996).

17. VandeVondele, J. & Hutter, J. Gaussian basis sets for accurate calculations on molecular systems in gas and condensed phases. *J. Chem. Phys.* **127**, 114105 (2007).

18. Kim, D. *et al.* Wettability of graphene and interfacial water structure. *Chem* **7**, 1602–1614 (2021).

19. Montenegro, A. *et al.* Asymmetric response of interfacial water to applied electric fields. *Nature* **594**, 62–65 (2021).

20. Wang, Y., Nagata, Y. & Bonn, M. Substrate effect on charging of electrified graphene/water interfaces. *Faraday Discuss.* (2023) doi:10.1039/D3FD00107E.

21. Bernhardt, N. *et al.* Large few-layer hexagonal boron nitride flakes for nonlinear optics. *Opt. Lett., OL* **46**, 564–567 (2021).

22. Kim, S. *et al.* Second-harmonic generation in multilayer hexagonal boron nitride flakes. *Opt. Lett., OL* **44**, 5792–5795 (2019).

23. Li, Y. *et al.* Probing symmetry properties of few-layer mos2 and h-bn by optical second-harmonic generation. *Nano Lett.* **13**, 3329–3333 (2013).

24. Moloney, E. G., Azam, Md. S., Cai, C. & Hore, D. K. Vibrational sum frequency spectroscopy of thin film interfaces. *Biointerphases* **17**, 051202 (2022).

25. Yamaguchi, S. (山口祥一), Shiratori, K. (白鳥和矢), Morita, A. (森田明弘) & Tahara, T. (田原太平). Electric quadrupole contribution to the nonresonant background of sum frequency generation at air/liquid interfaces. *J. Chem. Phys.* **134**, 184705 (2011).

26. Shen, Y. R. Revisiting the basic theory of sum-frequency generation. *J. Chem. Phys.* **153**, 180901 (2020).

27. Vandelli, M., Katsnelson, M. I. & Stepanov, E. A. Resonant optical second harmonic generation in graphene-based heterostructures. *Phys. Rev. B* **99**, 165432 (2019).

28. Yu, X., Chiang, K.-Y., Yu, C.-C., Bonn, M. & Nagata, Y. On the Fresnel factor correction of sum-frequency generation spectra of interfacial water. *J. Chem. Phys.* **158**, 044701 (2023).

29. Hale, G. M. & Querry, M. R. Optical constants of water in the 200-nm to 200-μm wavelength region. *Appl. Opt., AO* **12**, 555–563 (1973).

30. Wen, Y.-C. *et al.* Unveiling microscopic structures of charged water interfaces by surface-specific vibrational spectroscopy. *Phys. Rev. Lett.* **116**, 016101 (2016).

31. Ohno, P. E., Wang, H. & Geiger, F. M. Second-order spectral lineshapes from charged interfaces. *Nat. Commun.* **8**, 1032 (2017).

32. Reddy, S. K. *et al.* Bulk contributions modulate the sum-frequency generation spectra of water on model sea-spray aerosols. *Chem* **4**, 1629–1644 (2018).

33. Joutsuka, T. & Morita, A. Electrolyte and temperature effects on third-order susceptibility in sum-frequency generation spectroscopy of aqueous salt solutions. *J. Phys. Chem. C* **122**, 11407–11413 (2018).

34. Seki, T. *et al.* Real-time study of on-water chemistry: Surfactant monolayer-assisted growth of a

crystalline quasi-2D polymer. *Chem* **7**, 2758–2770 (2021).

35. Li, Z. *et al.* Ion transport and ultra-efficient osmotic power generation in boron nitride nanotube porins. *Sci. Adv.* **10**, eado8081 (2024).

36. Siria, A. *et al.* Giant osmotic energy conversion measured in a single transmembrane boron nitride nanotube. *Nature* **494**, 455–458 (2013).

37. R. Advincula, X. *et al.* Electrified/charged aqueous interfaces: general discussion. *Faraday Discuss.* **249**, 381–407 (2024).

38. Otero-de-la-Roza, A. & Johnson, E. R. Analysis of density-functional errors for noncovalent interactions between charged molecules. *J. Phys. Chem. A.* (2019) doi:10.1021/acs.jpca.9b10257.

39. Bryenton, K. R., Adeleke, A. A., Dale, S. G. & Johnson, E. R. Delocalization error: The greatest outstanding challenge in density-functional theory. *WIREs Comput. Mol. Sci.* **13**, e1631 (2023).